\documentclass[aps,prb,superscriptaddress,floatfix,twocolumn]{revtex4-2}

\usepackage[utf8]{inputenc}
\usepackage[T1]{fontenc}
\usepackage[english]{babel}
\usepackage{dsfont}
\usepackage{amsmath}
\usepackage{amsfonts}
\usepackage{amssymb}
\usepackage{graphicx}
\usepackage{hyperref}
\usepackage{physics}
\usepackage{bm}
\usepackage[separate-uncertainty]{siunitx}
\usepackage{gensymb}
\usepackage{wasysym}
\usepackage{xcolor}
\usepackage{xspace}
\newcommand{\moire}{\texorpdfstring{moir\'e\xspace}{moir\'e}}
\newcommand{\Moire}{\texorpdfstring{Moir\'e\xspace}{Moir\'e}}

\makeatletter
\renewcommand{\fnum@figure}{\textbf{Figure \thefigure}}
\renewcommand{\fnum@table}{\textbf{Table \thetable}}
\makeatother

\begin{document}
\title{Ultra-small-angle graphene twistronics with monolayer spacers: \\
From interlayer hybridisation and \moire effects \\
to high-temperature magnetotransport oscillations
}
\author{Benjamin Fuller}  
\email{benjamin.fuller@manchester.ac.uk}
\affiliation{Department of Physics and Astronomy, University of Manchester, Manchester, UK}
\affiliation{National Graphene Institute, University of Manchester, Manchester, UK}
\author{Haowei Fan}
\affiliation{Department of Physics, City University of Hong Kong, Kowloon, Hong Kong SAR}
\author{Agustin Montemurro}  
\affiliation{Department of Physics and Astronomy, University of Manchester, Manchester, UK}
\affiliation{National Graphene Institute, University of Manchester, Manchester, UK}
\author{Yuyuan Pan}
\affiliation{Department of Physics, City University of Hong Kong, Kowloon, Hong Kong SAR}
\author{Alexey I. Berdyugin}
\affiliation{Department of Materials Science and Engineering, National University of Singapore, Singapore, Singapore}
\affiliation{Department of Physics, National University of Singapore, Singapore, Singapore}
\affiliation{Institute for Functional Intelligent Materials,
National University of Singapore, Singapore, Singapore}
\author{Vladimir I. Fal'ko}  
\affiliation{Department of Physics and Astronomy, University of Manchester, Manchester, UK}
\affiliation{National Graphene Institute, University of Manchester, Manchester, UK}
\author{Xiao Li}
\email{xiao.li@cityu.edu.hk}
\affiliation{Department of Physics, City University of Hong Kong, Kowloon, Hong Kong SAR}

\begin{abstract}
Based on the analysis of \moire superlattice characteristics and  interlayer hybridisation in small-angle twisted graphene multilayers with a monolayer hexagonal boron nitride (hBN) spacer, we predict spectral features and
quantum transport effects in these systems. Here, we consider both a twisted bilayer and double bilayers with a spacer,  discuss double bilayers in two stacking orders---with and without inversion symmetry---and also compare the results obtained with self-consistent Hartree calculations with and without in-plane periodic charge distribution,
offering computationally efficient models for these graphene/hBN/graphene \moire systems.
We note parametric intervals where the appearance of minibands can be enhanced by tuning with an out-of-plane displacement field, 
and identify the miniband edges and van Hove singularities across a broad range of densities and displacement fields.
These system host Lifshitz transitions at which the Fermi lines form a network across \moire Brillouin zones, and to which Brown-Zak oscillations converge at low magnetic field and elevated temperatures.
We analyse interlayer hybridisation across the same parameter space, discuss how hybridisation affects in-plane magneto-transport in ultra-high-mobility structures, and predict that semiclassical magneto-oscillations of interlayer hybridisation can lead to $1/B$-periodic quantum oscillations of resistivity persistent to high temperature.
\end{abstract}

\maketitle

\section{Introduction}

\Moire systems with minibands present the opportunity to explore band edges and more tunable band structure at lower carrier concentrations and displacement fields than in untwisted monolayer or bilayer graphene~\cite{novoselovElectricFieldEffect2004,castronetoElectronicPropertiesGraphene2009,mccannLandauLevelDegeneracyQuantum2006,novoselovUnconventionalQuantumHall2006,mccannElectronicPropertiesBilayer2013,mccannAsymmetryGapElectronic2006}.
Twisted bilayer graphene (TBG) has been studied extensively in the context of flat bands, correlated phases and superconductivity~\cite{bistritzerMoireBandsTwisted2011,lopesdossantosGrapheneBilayerTwist2007,lopesdossantosContinuumModelTwisted2012,caoUnconventionalSuperconductivityMagicangle2018}.
A lot is known about strongly coupled TBG, however, there has been little research into weakly coupled \moire systems which could present interesting new opportunities.
While an atomically thin spacer between the graphene layers reduces the interlayer coupling, it enables
electrostatics to efficiently tune band alignment of the layers, adding an additional knob to control interlayer hybridisation and miniband formation in addition to the commonly-used `twist-angle' knob.
Here, for the spacer we consider a monolayer of hexagonal boron nitride (hBN), due to its atomically precise structure and large band gap
\cite{deanBoronNitrideSubstrates2010,xuGrapheneLikeTwoDimensionalMaterials2013,britnellElectronTunnelingUltrathin2012}.

In this study, we consider three exemplary twisted-with-a-spacer systems.
The first system is composed of two slightly misaligned monolayer graphenes separated by a monolayer of hBN, which we call ``MLG/hBN/MLG''.
Then we study two slightly misaligned Bernal bilayer graphenes separated by a monolayer of hBN.
These are analogous to twisted bilayer graphene (TBG) and twisted double bilayer graphene (TDBG), respectively, but with weaker interlayer coupling.
For the double bilayers there are two distinct cases: before twisting, the top bilayer may be parallel (P) or antiparallel (AP) to the bottom bilayer.
We call these systems ``parallel BLG/hBN/BLG'' and ``antiparallel BLG/hBN/BLG'' respectively, and we make this distinction because in a conventional TDBG without a spacer the relative stacking of the two bilayers produces a clear distinction in the band structure \cite{PhysRevB.99.235406}. 
Even a single monolayer of hBN is capable of suppressing the graphene-graphene coupling by more than an order of magnitude as compared to conventional twisted multilayers, scaling down relevant twist angles towards $\sim0.1\degree-0.5\degree$, 
simultaneously offering electrostatic tunability of moir{\'e} physics to new regimes.

The analysis performed in this work consists of the following steps:
\begin{itemize}
\item We introduce the single-particle model and Hartree mean-field corrections in Sec.~\ref{sec:model} and then calculate and present maps of DoS across a broad range of densities and displacement fields together with transport characteristics described in an effective scattering time approximation (see Sec.~\ref{sec:mlg} for MLG/hBN/MLG and Sec.~\ref{sec:blg} for BLG/hBN/BLG).
\item We compare parallel and antiparallel stackings of the two bilayers, and find only negligible differences in the DoS, hybridisation and transport characteristics in this weakly coupled system (Sec.~\ref{sec:blg}).
\item We compare $\bm{q}=0$ to $\bm{q}\ne 0$ terms in the self-consistent Hartree model and conclude that the minimal $\bm{q}=0$ model is sufficient to quantitatively describe all the miniband features in Sec.~\ref{subsec:Hartree discussion}
\item In the spirit of ref.~\cite{devriesKagomeQuantumOscillations2024}, we discuss the persistence of $1/B$-periodic Brown-Zak magneto-oscillations near Lifshitz transitions in the minibands at high temperatures and weak magnetic fields and map relevant parameter ranges of displacement field and carrier densities, see Sec.~\ref{subsec:BZ}.
\item We predict a peculiar type of quantum magneto-resistance oscillations 
 which sustain thermal broadening of the Fermi step.
These arise from magneto-oscillations of interlayer hybridisation between Dirac cones in the top and bottom graphenes which can be traced to the interference between pairs of tunnelling paths prescribed by momentum conservation in the interlayer hopping combined with cyclotron motion in the individual layers, see Sec.~\ref{subsec:hy}.
\end{itemize}

Furthermore, the separate effects of \moire minibands with umklapp scattering are contrasted with a toy model of tunnelling between two Dirac systems, in Appendix~\ref{subsec:toy}.
We studied a range of small twist angles for each system; results are shown in Sec.~\ref{sec:res}, and we illustrate the twist-angle dependence of \moire and hybridisation effects in Appendix~\ref{sec:other}.

\section{Model}\label{sec:model}

\begin{figure*}
  \includegraphics[width=\textwidth]{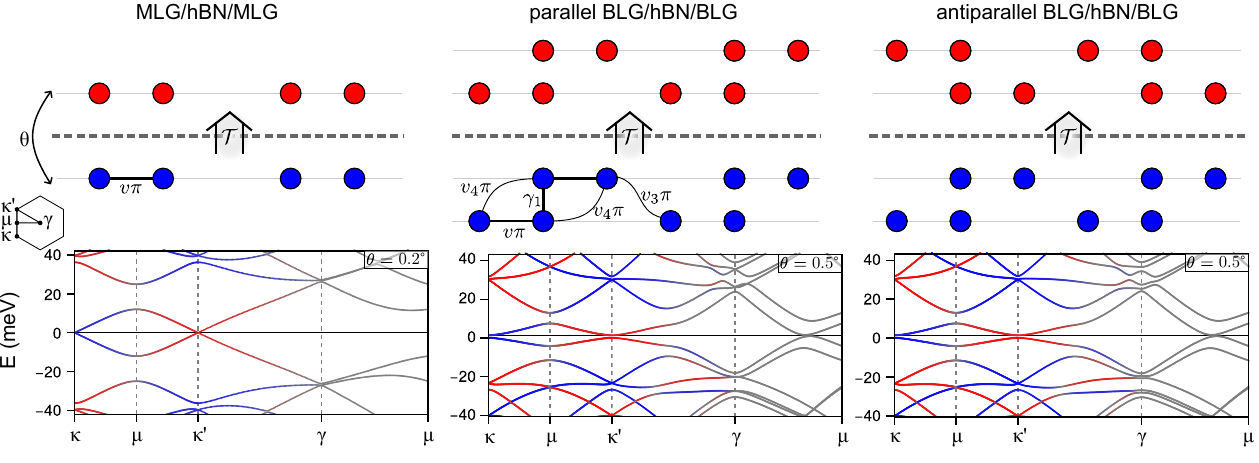} 
  \caption{Microscopic structures (top) and exemplary dispersions (bottom). Left: MLG/hBN/MLG, $0.2\degree$ twist. Center: parallel BLG/hBN/BLG, $0.5\degree$ twist. Right: antiparallel BLG/hBN/BLG, $0.5\degree$ twist. Dashed gray line is the hBN spacer. The dispersions are calculated at 0 filling and $D=0$ and the solid black line is the Fermi level at charge neutrality. The vertical axis is energy in meV. Red/blue colour indicates carrier polarisation $\Delta(\bm{k},n)$ on top/bottom layer(s), defined in Eq.~\eqref{eq:polarisation}. Inset shows the high-symmetry path followed around the mBZ for the dispersions.}
  \label{fig1}
\end{figure*}

  Firstly, we present the MLG/hBN/MLG model in detail.
  The treatment of systems of double bilayers is very similar and is discussed below.
  The microscopic structure of the systems are shown in Fig.~\ref{fig1} along with exemplary dispersions showing minibands.

\subsection{Single-particle Hamiltonian}

We describe the systems using a hybrid $k\cdot p$-theory--tight-binding (HkpTB) approach.
The single-particle Hamiltonian in one valley is
  \begin{equation}
  \label{eq:H}
      \mathcal{H}_\text{sp} = \begin{pmatrix}
          H_b & \mathcal{T}\\
          \mathcal{T}^\dag & H_t
      \end{pmatrix}, 
  \end{equation}
with the Hamiltonian for monolayer graphene
\begin{equation}
  H_{b/t} = \hbar v \bm{\sigma}^\xi \cdot (\bm{k}-\bm{K}_{b/t}) + \epsilon_{b/t} \openone, 
\end{equation}
where the index $b/t$ indicates the bottom or top layer, $\epsilon_{b/t}$ are onlayer potentials discussed in the next subsection, $\hbar$ is the reduced Planck constant, $\xi=\pm 1$ is the valley index, $\bm{\sigma}^\xi = (\xi \sigma_x, \sigma_y)$ is a vector of Pauli matrices in the (A,B) sublattice basis, $v$ is the Fermi velocity, wavevector $\bm{k}$ is measured from the rotated Dirac points
$\bm{K}_b = R(-\theta/2)\bm{K}_\xi^{(0)}$, $\bm{K}_t = R(\theta/2)\bm{K}_\xi^{(0)}$,
Dirac points are $\bm{K}^{(j)}_\xi=\xi\frac{4\pi}{3a}(\cos(2\pi j/3), -\sin(2\pi j/3))$,
$j=0,1,2$ accounts for the three equivalent corners of the Brillouin zone,
$a$ is the lattice constant of graphene, and $\theta$ is the twist angle between top and bottom layers.
The \moire interlayer tunnelling matrix is
\begin{equation}
\label{eq:T}
  \mathcal{T}
      = \gamma \sum_{j=0}^2
      \begin{pmatrix}
        1 & e^{i \xi 2\pi j/3} \\
        e^{-i\xi 2\pi j/3} & 1
      \end{pmatrix} \delta_{\bm{k}'\!,\,\bm{k}+\bm{\Delta K}^{(0)}_\xi-\bm{\Delta K}^{(j)}_\xi},
\end{equation}
where $\Delta \bm{K}^{(j)}_\xi=\theta\hat{\bm{z}}\times \bm{K}^{(j)}_\xi$ is the shift as a result of the small twist $\theta\ll 1$ between Dirac points \cite{garcia-ruizFullSlonczewskiWeissMcClureParametrization2021}, which gives an additional periodicity in the system and causes the formation of minibands in \moire Brillouin zones (mBZs).
This single-particle model is analogous to the Bistritzer-MacDonald model for TBG \cite{bistritzerMoireBandsTwisted2011,lopesdossantosContinuumModelTwisted2012,suarezmorellFlatBandsSlightly2010,meleCommensurationInterlayerCoherence2010,guineaContinuumModelsTwisted2019a}, but with a modified interlayer tunnelling strength $\gamma$ to account for the presence of the spacer.
The twisted interface is considered in a continuum approximation:
we construct a basis of states in which the \moire potential couples momenta which differ by a \moire Bragg vector $\bm{G}$.

The spacer affects the interlayer coupling strength $\gamma$.
The interlayer coupling strength is the most important parameter; the ratio of interlayer coupling to twist angle determines the strength or relevance of \moire effects.
It is estimated by taking the average of hoppings past boron or nitrogen sites, which gives a strength $\gamma\approx\SI{7}{meV}$ \cite{chenSemimetallicSemiconductingGraphenehBN2023}.
This is about $15$ times weaker than in systems where the twisted layers are in direct contact~\cite{PhysRevB.99.235406}.
This means that in order for interlayer hybridisation to have a noticeable effect on the electronic dispersion, the twist angle $\theta$ between layers must be correspondingly smaller.
In the main text we highlight $\theta=\SI{0.2}{\degree}$ for the MLG/hBN/MLG system in detailed numerical calculations.
This ratio of twist angle to interlayer coupling is equivalent to a $\SI{3}{\degree}$ twist in conventional TBG without a spacer. 
For the BLG/hBN/BLG systems (both ABAB and ABBA stacking orders), we highlight $\theta=0.5\degree$ which hosts minibands at comparable energies to the monolayer system, as shown in Fig.~\ref{fig1}.
These configurations are experimentally achievable \cite{yooAtomicElectronicReconstruction2019}, show some \moire effects, and are easily tunable by an appropriate range of displacement field.
Results for other twist angles are given in Appendix~\ref{sec:other}.

We also study the effect of an hBN spacer between bilayers of graphene, like twisted double bilayer graphene (TDBG).
Each bilayer is Bernal-stacked, and there are two configurations for the orientation of each bilayer: parallel (AB/AB or BA/BA) and antiparallel (AB/BA or BA/AB).
To adapt the model presented above for twisted double bilayers with a spacer, the basis is twice as large (with 4 layers) and the single particle Hamiltonian for one AB Bernal bilayer graphene is, in the basis (A1, B1, A2, B2) \cite{mccannLandauLevelDegeneracyQuantum2006,mccannElectronicPropertiesBilayer2013,PhysRevB.99.235406},
\begin{equation}
  H^{AB}_{b/t}(\bm{K}_{b/t}) =
  \begin{pmatrix}
    \epsilon_{1/3} & v \pi^\dag & -v_4\pi^\dag & -v_3 \pi\\
    v \pi  & \epsilon_{1/3} + \Delta' & \gamma_1 & -v_4 \pi^\dag\\
    -v_4 \pi  & \gamma_1  & \epsilon_{2/4} + \Delta' & v \pi^\dag \\
    -v_3\pi & -v_4 \pi & v \pi & \epsilon_{2/4}
  \end{pmatrix},
\end{equation}
and the BA stacking is obtained by rotating this $180\degree$, or by switching A and B sublattices:
\begin{equation}
  H^{BA}_{b/t}(\bm{K}_{b/t}) =
  \begin{pmatrix}
    \epsilon_{1/3} + \Delta' & v \pi^\dag & -v_4\pi & \gamma_1\\
    v \pi  & \epsilon_{1/3} & -v_3 \pi^\dag & -v_4 \pi\\
    -v_4\pi^\dag    &-v_3 \pi & \epsilon_{2/4} & v \pi^\dag \\
    \gamma_1  & -v_4 \pi^\dag & v \pi & \epsilon_{2/4} + \Delta'
  \end{pmatrix},
\end{equation}
where $\pi = \hbar (\xi(k_x-K_x) + i (k_y-K_y))$,
$\Delta'$ is the dimer site energy shift compared to nondimer sites,
$\gamma_1$ is the intrabilayer coupling between dimer sites, $v_3$ is a diagonal hopping between nondimer sites, and $v_4$ is a diagonal hopping between one dimer and one nondimer site.
Since there are more layers in this system, we enumerate the onlayer potentials $\epsilon_l$ with $l=1,2,3,4$, where layers 1 and 2 form the bottom ($b$) bilayer and layers 3 and 4 form the top ($t$) bilayer.
So, for the parallel configuration, we use $H^{AB}_b$ and $H^{AB}_t$, whereas for the antiparallel configuration we use $H^{AB}_b$ and $H^{BA}_t$.
The \moire coupling is only between the innermost layers, so the interlayer coupling is described by replacing Eq.~\eqref{eq:T} with the matrix form as
\begin{equation}
\mathcal{T} \to
\begin{pmatrix}
  0 & 0\\
  \mathcal{T} & 0
\end{pmatrix}.
\end{equation}
The rest of the model is equivalent to the model for MLG/hBN/MLG but with the new basis.

We also considered using effective models for each bilayer.
It is well-known that the dimer pair of orbitals hybridise to a high energy band, leaving the non-dimer orbitals dominating the low-energy dispersion, which permits to write a computationally simpler $2\times2$ model for the non-dimer sites
\cite{mccannLandauLevelDegeneracyQuantum2006,mccannElectronicPropertiesBilayer2013}.
However, the accuracy of this low-energy model is reduced when the dimer and non-dimer bands are mixed as a result of high displacement field or by probing higher fillings.
Therefore, the results presented here are obtained using the full $4\times4$ model for each bilayer.

\subsection{Self-consistent Hartree analysis}\label{subsec:Hartree approximation}

The full model includes single-particle and interacting parts, where the interaction is introduced by the Hartree term $\mathcal{H}_\text{full} = \mathcal{H}_{\text{sp}}+V_\text{H}^t + V_\text{H}^b$~\cite{guineaElectrostatics2018,kwanMeanfieldModelingMoire2025,ceaElectronicBandStructure2019,ceaElectrostaticInteractionsTwisted2022,ceaBandStructureInsulating2020,kwanMoireFractionalChern2025,rademakerChargeSmootheningBand2019}.
We separate the full Hartree interaction into $\bm{q}=0$ and $\bm{q}\ne 0$ parts,
\begin{equation}
  V_\text{H}^l = V_0^l + \sum_{\bm{q}\ne0} \left(V_{\bm{q}}^{l,l} + V_{\bm{q}}^{l,l'}\right),
\end{equation}
where $\bm{q}$ is the momentum transfer in the interaction.
The first term, $\bm{q}=0$ part, is strongest.
It sets homogeneous onlayer potentials $\epsilon_l$:
\begin{equation}
  V_0^l = \epsilon_l \sum_{\bm{k},s} \hat{c}^\dag_{\bm{k},l,s}\hat{c}_{\bm{k},l,s},
  \label{eq:onlayer}
\end{equation}
where $\hat{c}_{\bm{k},l,s}$ is an annihilation operator for the state at momentum $\bm{k}$, layer $l$, and sublattice $s=A,B$.
The layer index $l$ takes the values $t,b$ for MLG/hBN/MLG or 1,2,3,4 for BLG/hBN/BLG.
The $\bm{q}\ne 0$ contribution describes in-plane charge fluctuation with the periodicity of the superlattice, and it is much weaker.
It is discussed below.
We neglect the screening of the Coulomb potential by external gates, assuming that the distance to the gates $d$ is much longer than the periodicity of the \moire unit cell, $L_M=a/\theta$.
We adopt the infinite-temperature reference field for the Hartree interaction~\cite{kwanMoireFractionalChern2025,huangSelfconsistentTheoryFractional2024}.
The explicit construction and the associated self-consistent procedure are discussed below.
The Fock (exchange) interaction is neglected here because of the weak interlayer coupling.
Moreover, ref.~\cite{ceaElectronicBandStructure2019} showed that the exchange interaction only makes small quantitative changes to the DoS without qualitative changes of features.

The hBN spacer has two effects: it increases the distance between the graphene layers, and it decreases the tunnelling strength between them.
The increased distance gives the external displacement field a larger distance to act over, thereby enhancing the induced interlayer potential difference.
The onlayer potentials for the MLG/hBN/MLG model are found by calculating the voltage drop between each layer of graphene
\cite{slizovskiyOutofPlaneDielectricSusceptibility2021}:
\begin{align}
  \epsilon_t - \epsilon_{b} = 
  &\frac{eD}{\varepsilon_0}\qty(\frac{d_\text{G}}{\varepsilon_z^\text{G}} + \frac{d_\text{hBN}}{\varepsilon_z^\text{hBN}}) \nonumber\\
  &+ \frac{e^2 }{\varepsilon_0}\qty[\frac{1+\varepsilon_z^\text{G}}{2}\frac{d_\text{G}}{\varepsilon_z^\text{G}}+  \frac{d_\text{hBN}}{\varepsilon_z^\text{hBN}}]\frac{(n_t-n_b)}{2},
  \label{eq:u}
\end{align}
where
$n_l$ is the carrier density on layer $l$,
$d_\text{G}$ and $d_\text{hBN}$ are the widths of monolayer graphene and hBN, and
$\varepsilon_z^\text{G}$ and $\varepsilon_z^\text{hBN}$ are the out-of-plane dielectric permittivities of graphene and hBN \cite{slizovskiyOutofPlaneDielectricSusceptibility2021, laturia2018dielectric,pierret2022dielectric,wehlingStrengthEffectiveCoulomb2011}.
This equation was derived from ref.~\cite{slizovskiyOutofPlaneDielectricSusceptibility2021} by adding the voltage drop from the hBN dielectric but keeping the screening correction due to polarisability of graphene only on the graphene parts.

For the BLG/hBN/BLG model, the onlayer potentials are similarly determined as follows \cite{slizovskiyOutofPlaneDielectricSusceptibility2021,garcia-ruizMixedStackingFewLayerGraphene2023}
\begin{widetext}
\begin{equation}
  \begin{split}
  \epsilon_2 - \epsilon_{1} &= 
  \frac{eD}{\varepsilon_0}\qty(\frac{d_\text{G}}{\varepsilon_z^\text{G}}) + \frac{e^2 }{\varepsilon_0}\qty[\frac{1+\varepsilon_z^\text{G}}{2}\frac{d_\text{G}}{\varepsilon_z^\text{G}}]\frac{(n_2-n_{1})}{2} + \frac{e^2}{2\varepsilon_0} \qty(\frac{d_\text{G}}{\varepsilon_z^\text{G}}) \qty(n_3+n_4), \\
  \epsilon_3 - \epsilon_{2} &= 
  \frac{eD}{\varepsilon_0}\qty(\frac{d_\text{G}}{\varepsilon_z^\text{G}} + \frac{d_\text{hBN}}{\varepsilon_z^\text{hBN}}) + \frac{e^2 }{\varepsilon_0}\qty[\frac{1+\varepsilon_z^\text{G}}{2}\frac{d_\text{G}}{\varepsilon_z^\text{G}}+  \frac{d_\text{hBN}}{\varepsilon_z^\text{hBN}}]\frac{(n_3-n_{2})}{2} + \frac{e^2}{2\varepsilon_0} \qty(\frac{d_\text{G}}{\varepsilon_z^\text{G}}+\frac{d_\text{hBN}}{\varepsilon_z^\text{hBN}}) \qty(n_4 - n_1), \\
  \epsilon_4 - \epsilon_{3} &= 
  \frac{eD}{\varepsilon_0}\qty(\frac{d_\text{G}}{\varepsilon_z^\text{G}}) + \frac{e^2 }{\varepsilon_0}\qty[\frac{1+\varepsilon_z^\text{G}}{2}\frac{d_\text{G}}{\varepsilon_z^\text{G}}]\frac{(n_4-n_{3})}{2} - \frac{e^2}{2\varepsilon_0} \qty(\frac{d_\text{G}}{\varepsilon_z^\text{G}}) \qty(n_1+n_2).
  \end{split}
\end{equation}
\end{widetext}

In addition to the homogeneous onlayer potential described in Eq.~\eqref{eq:onlayer}, there is an additional charge density fluctuation due to the superlattice potential.
This $\bm{q}\ne 0$ contribution arises from the in plane charge density modulation due to the superlattice. Therefore, the charge density fluctuation is nonzero for momentum transfer equal to a \moire Bragg vector, $\bm{q} = \bm{G}$.
This interaction introduces scattering between momentum states in the same layer and sublattice 
\begin{equation}\label{eq:vq}
\begin{split}
   V_{\bm{q}}^{l,l} + V_{\bm{q}}^{l,l'} = \frac{1}{A} \frac{e^2}{2\varepsilon_0\varepsilon_\parallel q} 
   \Big(
     \expval{\delta\rho_{-\bm{q},\,l}}\hat{\rho}_{\bm{q},l}  \\
     + \expval{\delta\rho_{-\bm{q},\,l'}}\hat{\rho}_{\bm{q},l}\ e^{-qd} \Big)
\end{split}
\end{equation}
for (bi)layer $l=t,b$ and $l' \ne l$,
where $q=|\bm{q}|$, $A$ is the total system real area which is related to the k-space grid density by $A=N_k A_{\varhexagon}$ for $N_k$ points sampled across the mBZ and real area of the \moire unit cell $A_{\varhexagon}$, $\varepsilon_\parallel$ is the in-plane dielectric constant of the environment, $d$ is the interlayer distance,
and the density operator in $k$-space (or Fourier transform of density operator) is
\begin{equation}
  \hat{\rho}_{\bm{q},\,l} = \frac{4}{N_k} \sum_{\bm{k},\bm{G},s} c^\dag_{\bm{k}+\bm{G}+\bm{q},l,s} c_{\bm{k}+\bm{G},l,s},
\end{equation}
where
$\bm{k}$ ranges over occupied states in the first \moire Brillouin zone (mBZ),
4 accounts for spin and valley degeneracy,
$N_k$ is the number of k-points sampled across the mBZ,
and $\bm{G}$ are \moire Bragg vectors (within a finite cutoff $G_m$).

According to Eq.~\eqref{eq:vq}, the scattering amplitude decreases with higher momentum transfer, so the first shell is most important and higher-order shells have negligible effect for our systems.
The averaged density $\expval{\delta\rho_{\bm{q},l}}$ is summed over all occupied states by strictly picking states with eigenenergy below the Fermi level.
The fluctuation density (relative to charge neutrality) is given by $\delta\hat{\rho}_{\bm{q}, \, l} = \hat{\rho}_{\bm{q}, \, l} - \expval{\hat{\rho}_{\bm{q}, \, l}}_{\text{ref}}$. We take the infinite-temperature reference field $\expval{\hat{\rho}_{\bm{q}, \, l}}_{\text{ref}}=\openone/2$ \cite{huangSelfconsistentTheoryFractional2024}.
The ground state Hartree potential is found by a self-consistent iteration:
we begin with an initial state $\delta\hat{\rho}_{\bm{q},l}=0$ and numerically solve the Hamiltonian across a grid in the mBZ.
From the calculated eigenenergies and eigenvectors, we obtain the density matrix $\delta\hat\rho^\text{(new)}$.
Then we perform a fixed-point iteration to find the next density matrix from an interpolation of the previous and new calculated value, $\delta\hat{\rho}^{(n+1)} = (1-\eta)\delta\hat{\rho}^{(n)} + \eta\, \delta\hat\rho^\text{(new)}$ where $\eta$ is determined by the optimal damping algorithm \cite{kwanMeanfieldModelingMoire2025}.

To describe the in-plane dielectric constant of the environment we choose $\varepsilon_\parallel=7$, but we note that it varies over a wide range of values in the literature, around 4--10 \cite{ceaElectronicBandStructure2019}.
We neglect the intervalley Coulomb interaction and only model the intravalley part. In fact, we neglect all intervalley scattering, because the twist angles are very small, and calculate everything for each valley independently.

\section{Results}\label{sec:res}

With the continuum \moire minibands and Hartree interaction we model the behaviour of the system across a range of fillings and accounting for an external out-of-plane displacement field.
Specifically, we calculate density of states at the Fermi level (DoS), interlayer hybridisation, and longitudinal resistivity.
We also calculate the Hall resistivity in a weak magnetic field.

\subsection{MLG/\lowercase{h}BN/MLG}
\label{sec:mlg}

\begin{figure*}
    \includegraphics[width=\textwidth]{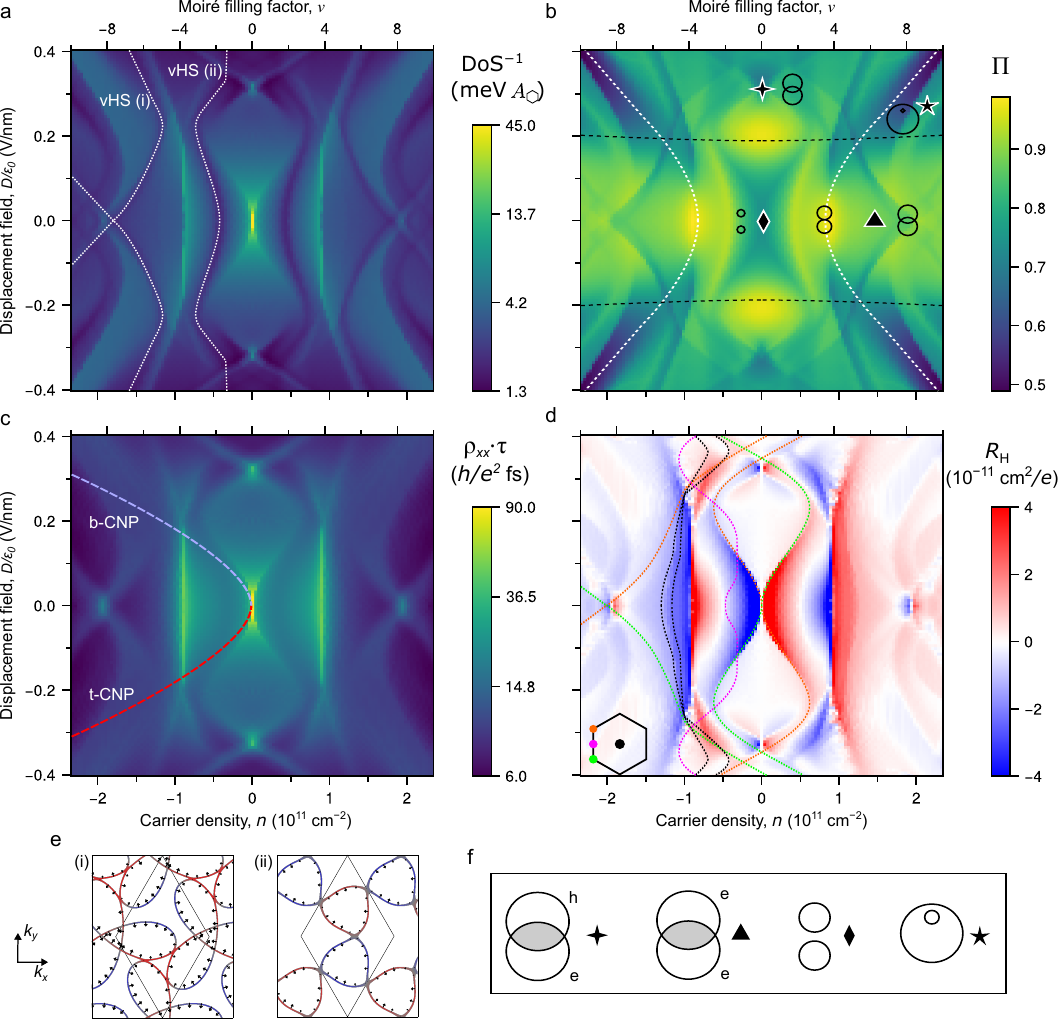}
    \caption{
    Results for MLG/hBN/MLG, $\theta=0.2\degree$. Maps are for varying external displacement field $D$ against carrier density $n$ (bottom ticks), which corresponds to filling factor of the \moire unit cell $\nu$ (top ticks).
    \textbf{(a)}: inverse density of states at the Fermi level.
    Dotted lines trace positions of Lifshitz transitions due to van Hove singularities (vHS), of two types (i) and (ii) for which exemplary Fermi contours are shown in (e).
    The right halves of the maps are equivalent, just with layers switched, so for clarity they are not annotated with lines.
    \textbf{(b)}: hybridisation between top and bottom layers, $\Pi$, defined in Eq.~\eqref{eq:pi}.
    White dashed lines shows where the Fermi level lies at the threshold of Fermi circles touching.
    Horizontal black dashed lines show the critical displacement field $D^*$ where cones coincide along a line at all densities.
    The polygons label the distinct regions depending on the Fermi contours as shown in (f):
    rhombus and 5-star show non-intersecting (disjoint) circles,
    4-star is intersecting circles of opposite carrier type,
    triangle is intersecting circles of same carrier type.
    \textbf{(c)}: longitudinal resistivity normalised by scattering time, $\rho_{xx}\cdot\tau$.
    Dashed lines show positions of the charge neutrality points (CNPs) for bottom and top layers in blue and red.
    \textbf{(d)}: Hall coefficient.
    Lines depict where high-symmetry momentum states are occupied at the Fermi level, according to the inset: black for $\gamma$, green/orange for $\kappa$/$\kappa'$, pink for $\mu$.
    \textbf{(e)}: Fermi contours showing two instances of van Hove singularities.
    Red/blue colour indicates carrier polarisation on top/bottom layer, $\Delta(\bm{k},n)$, according to Eq.~\eqref{eq:polarisation}.
    Arrows show the direction and magnitude of band velocity. The rhombus in the background outlines one \moire Brillouin zone, centred at $\mu$.
    \textbf{(f)}: the possible arrangements of Fermi circles of each monolayer graphene, each corresponding to a region in (b).
    }
    \label{fig:res1}
\end{figure*}

The main results for MLG/hBN/MLG system are shown in Fig.~\ref{fig:res1} for $\theta=0.2\degree$ in maps for varying total carrier density $n$ and external out-of-plane displacement field $D$.
The top axis of the maps shows carrier density normalised by the \moire unit cell area to give filling factors $\nu=n/n_0;\ n_0=1/ A_{\varhexagon}$.
To develop an understanding of the system we inspect the map of inverse DoS.
The map shows many features across a small range of displacement field and filling.
Bright regions are low DoS and would correspond to high resistance in transport measurements.
Some of the darkest regions are identified as corresponding to a Lifshitz transition (LT) due to a van Hove singularity (vHS) at the Fermi level.
These are shown by dotted lines for two types of vHS, and an exemplary Fermi contour of such points is given in panel (e).

To understand which effects are responsible for various features in the maps, we compared the results of three models: (i) graphene layers separated by hBN with no interlayer tunnelling, only interacting electrostatically; (ii) a toy model of two 2D Dirac cones that are offset in momentum by $\Delta K$ and can hybridise, but without umklapp scattering processes that give rise to minibands or mini Brillouin zone periodicity; (iii) the full \moire model which includes electrostatics and umklapp scattering processes.
Model (i) is obtained by setting the interlayer tunnelling parameter $\gamma$ to 0.
The toy model (ii) is obtained by selecting one of the three tunnelling terms in Eq.~\eqref{eq:T}.
This direct interlayer hybridisation without umklapp scattering is representative of the tunnelling measurements performed in devices with thicker spacers \cite{mishchenkoTwistcontrolledResonantTunnelling2014}, or in the quantum twist microscope \cite{inbar2023quantum}.
The results of each individual model are shown in the Appendix, Fig.~\ref{fig:strip}.
With this analysis, the origins of each feature can be understood.
Namely, the electrostatics produces the charge neutrality points (CNPs) of individual layers which appear as a parabolic horizontal `X' shape out from the centre.
Hybridisation without umklapp generates a new ``stripe'' for the valence and conduction sides, where the Dirac cones of the same carrier type cross and hybridise.
Therefore, the other details of the full model that are missing from the results of (i) and (ii) are a signature of \moire physics, particularly the vertical features at 4$\times$integer fillings of the \moire unit cell, where the factor 4 is for spin and valley degeneracy.

To quantify the hybridisation between top and bottom layers or bilayers, we
use the product of projected weight onto bottom and top layers, and define a measure of hybridisation
\begin{equation}
\label{eq:pi}
    \Pi = 2\frac {\sum_{n,\bm{k}} -\frac{\partial f_0(\epsilon_{n,\bm{k}})}{\partial \epsilon}
    \sqrt{
    \textstyle\sum\limits_{s} |\psi_n^{bs}(\bm{k})|^2
    \textstyle\sum\limits_{s} |\psi_n^{ts}(\bm{k})|^2
    }}{\sum_{n,\bm{k}} -\frac{\partial f_0(\epsilon_{n,\bm{k}})}{\partial \epsilon}},
\end{equation}
where $\psi_n^{ls}(\bm{k})$ indicates the wave functions for band $n$ at $\bm{k}$ hosted on layer $l$ and sublattice $s=A,B$.
$\Pi$ indicates hybridisation: $\Pi=1$ indicates maximally mixed states with delocalised carriers, and $\Pi=0$ indicates pure (unmixed) states with layer polarised carriers.
The red/blue colour of lines in the dispersions in Fig.~\ref{fig1} and Fermi contours in Fig.~\ref{fig:res1}(e) are determined by the carrier top/bottom polarisation
\begin{equation}
    \Delta(\bm{k}, n) = \sum_s |\psi_n^{ts}(\bm{k})|^2 - \sum_s |\psi_n^{bs}(\bm{k})|^2,
    \label{eq:polarisation}
\end{equation}
which lies between $-1$ and 1, indicating carriers polarised on the top or bottom layers.

The hybridisation $\Pi$ is plotted in Fig.~\ref{fig:res1}b.
The map shows a wide range of hybridisation.
Maximum hybridisation occurs in two main regions when bands from the top and bottom layers touch: at $|D|/\varepsilon_0=\SI{0.2}{V/nm}$ and low density, where the critical displacement field causes the monolayer Dirac cones to coincide along a line by setting the difference in onlayer potentials $u=\hbar v\Delta K$, enabling resonant hybridisation \cite{britnellResonantTunnellingNegative2013,mishchenkoTwistcontrolledResonantTunnelling2014,laneTwistcontrolledResonantTunnelling2015};
and secondly at $|\nu|\approx4$ and $D=0$, where the Fermi contours in each layer have grown until they touch.
The points where the cones just touch are noted by the white dashed lines, and the resonant tunnelling condition is shown by the horizontal black dashed lines.
These lines divide the map into distinct regions: 
where Fermi contours are disjoint and do not touch (centre and corners); where Fermi contours of the same type intersect (left and right sides); and where Fermi contours of opposite carrier type intersect (top and bottom).
These are marked with the polygons and illustrated by circles that are cartoons of the monolayer Fermi contours, on the map and corresponding in panel (f).

The classical contribution to resistivity is found by calculating band velocities and effective masses.
The expressions used are given in Appendix~\ref{Conductivity calculation}, where we employ the scattering-time approximation and assume a constant scattering time $\tau$ independent of wavevector which characterises any scattering processes.
The longitudinal resistivity is shown in panel (c), normalised by scattering time.
It is a measure of band velocity multiplied by DoS, and it shows similar features to the DoS map.
The Hall coefficient is shown in panel (d).
We note multiple sign changes when changing fillings by integer $\nu$, which is a signature of miniband formation.
The dotted lines on panel (d) show points where high-symmetry points of the mBZ are occupied at the Fermi level.
These high-symmetry points are usually occupied at band edges or saddle points.

Similar twisted structures with a spacer have been studied at slightly larger twist angles \cite{mishchenkoTwistcontrolledResonantTunnelling2014,laneTwistcontrolledResonantTunnelling2015}, where the resonant hybridisation effect was noted and the tunability by an electric field.
In this work we show that \moire physics can be added to the system by exploring smaller twist angles.
In Appendix~\ref{sec:other}, we show that \moire features become noticeable only at or below a small twist angle of about $0.7\degree$, but probing much smaller twist angles increases the \moire strength such that the carrier hybridisation is less tunable.

\subsection{BLG/\lowercase{h}BN/BLG}
\label{sec:blg}

\begin{figure*}
    \includegraphics[width=\textwidth]{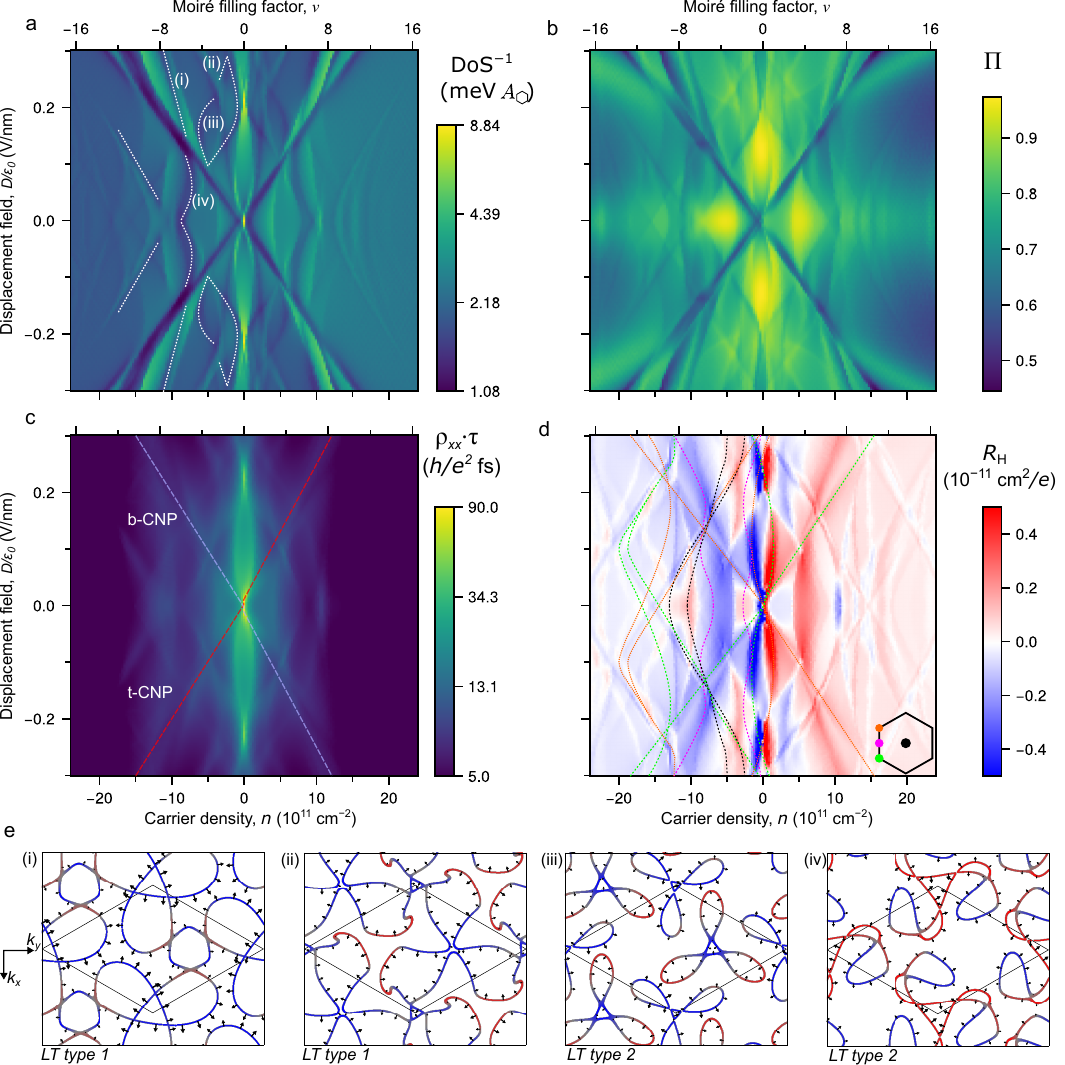}
    \caption{Results for parallel (inversion symmetric) BLG/hBN/BLG, $\theta=0.5\degree$. Maps are for varying external displacement field $D$ against carrier density $n$ (bottom ticks), which corresponds to filling factor of the \moire unit cell $\nu$ (top ticks).
    \textbf{(a)}: inverse density of states at the Fermi level. Dotted lines trace positions of Lifshitz transitions (LTs) due to van Hove singularities, some of which are shown in (e).
    The right halves of the maps are equivalent so are not annotated with lines for clarity.
    \textbf{(b)}: hybridisation $\Pi$ between top and bottom bilayers, defined in Eq.~\eqref{eq:pi2}.
    \textbf{(c)}: longitudinal resistivity normalised by scattering time, $\rho_{xx}\cdot\tau$ . Dashed lines show positions of the CNPs for bottom and top layers in blue and red.
    \textbf{(d)}: Hall coefficient.
    Dotted lines depict where high-symmetry momentum states are occupied at the Fermi level, according to the inset: black for $\gamma$, green/orange for $\kappa/\kappa'$, pink for $\mu$.
    \textbf{(e)}: Fermi contours showing various Lifshitz transitions of two types corresponding to dotted lines in (a).
    Red/blue colour indicates carrier polarisation on top/bottom bilayer, $\Delta(k, n)$, according to Eq.~\eqref{eq:polarisation2}.
    Arrows show the direction and magnitude of band velocity.
    The rhombus in the background outlines one \moire Brillouin zone centred at $\mu$.
    }
    \label{fig:res2p}
\end{figure*}

\begin{figure*}
    \includegraphics[width=\textwidth]{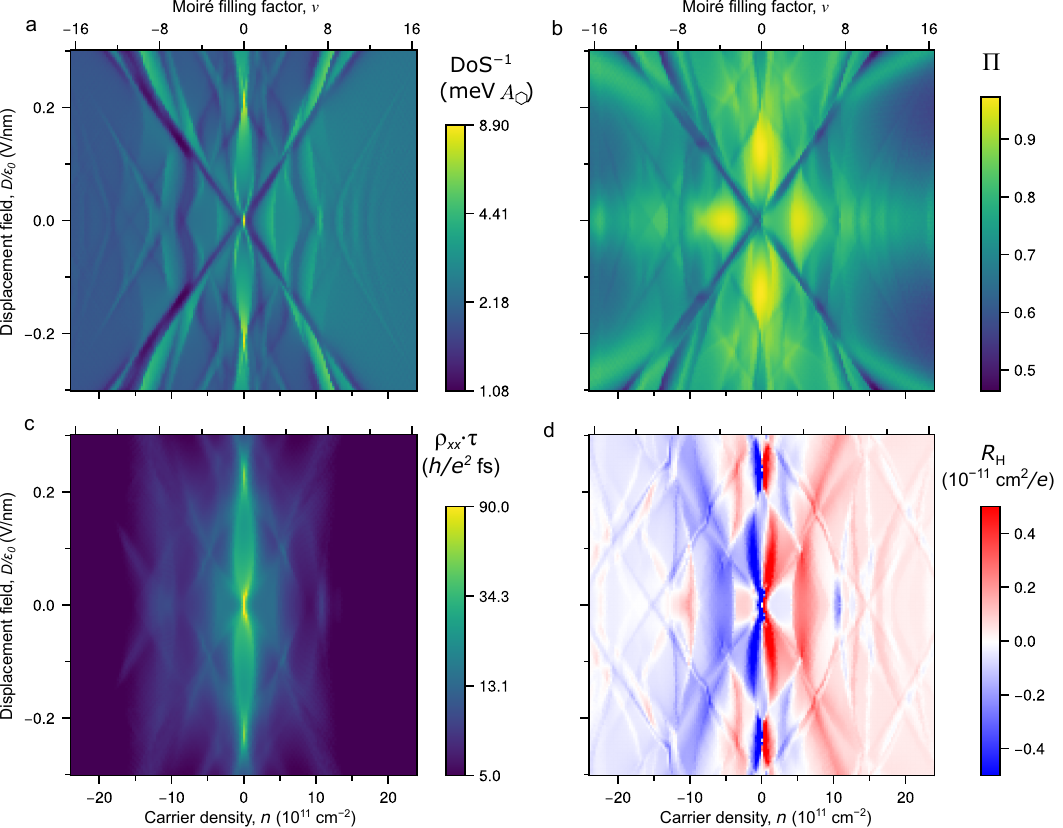}
    \caption{Results for antiparallel (inversion asymmetric) BLG/hBN/BLG, $\theta=0.5\degree$. Maps are for varying external displacement field $D$ against carrier density $n$ (bottom ticks), which corresponds to filling factor of the \moire unit cell $\nu$ (top ticks).
    \textbf{(a)}: inverse density of states at the Fermi level.
    \textbf{(b)}: hybridisation $\Pi$ between top and bottom bilayers, defined in Eq.~\eqref{eq:pi2}.
    \textbf{(c)}: longitudinal resistivity normalised by scattering time, $\rho_{xx}\cdot\tau$.
    \textbf{(d)}: Hall coefficient.
    }
    \label{fig:res2ap}
\end{figure*}

The main results for the BLG/hBN/BLG systems are shown in Fig.~\ref{fig:res2p} for parallel (AB/AB) and Fig.~\ref{fig:res2ap} for antiparallel (AB/BA) stackings, both at $\theta=0.5\degree$.
The maps show many features across a modest range of displacement field and filling.
Bright regions represent low DoS and would correspond to high resistance in transport measurements.
The most notable feature is a prominent `X' shape where hybridisation $\Pi$ is low, and at the edges of which there is a high DoS.
This occurs when the Fermi level lies in a gap of one of the bilayers.
Then, all the carriers are hosted on the other bilayer only.
At the sides of this gap are band edges, which are relatively flat and host high DoS, resulting in the dark lines in panel (a).
The dashed red and blue lines on panel (a) indicate the Fermi level aligning with CNPs of top or bottom bilayer, which corresponds with this gap.
Such a gap is not possible in the monolayers system.

Analogously with the monolayers system we plot a measure of hybridisation
\begin{equation}
    \Pi =  2\frac{\sum_{n,\bm{k}} -\frac{\partial f_0(\epsilon_{n,\bm{k}})}{\partial \epsilon}
    \sqrt{
    \textstyle\sum\limits_{s,l=1,2} |\psi_n^{ls}(\bm{k})|^2
    \textstyle\sum\limits_{s,l=3,4} |\psi_n^{ls}(\bm{k})|^2
    }}{\sum_{n,\bm{k}} -\frac{\partial f_0(\epsilon_{n,\bm{k}})}{\partial \epsilon}}
    \label{eq:pi2}
\end{equation}
and top/bottom polarisation
\begin{equation}
    \Delta(\bm{k}, n) = \sum_{s,l=3,4} |\psi_n^{ls}(\bm{k})|^2 - \sum_{s,l=1,2} |\psi_n^{ls}(\bm{k})|^2.
    \label{eq:polarisation2}
\end{equation}

We check both stacking orders because
in a conventional twisted double bilayer without a spacer the relative stacking of the two bilayers produces a clear distinction in the band structure \cite{PhysRevB.99.235406}.
Here, however, the hBN spacer both suppresses the absolute magnitude of the interlayer coupling and increases the spatial separation between the two bilayers.
Consequently the difference between the two stacking orders is strongly reduced, and the DoS, hybridisation and semiclassical transport maps of the two geometries are essentially indistinguishable within the resolution of our maps.

To understand the origin of the features shown in the inverse DoS in panel (a), we performed the same procedure as for the MLG/hBN/MLG system, shown in the Appendix Fig.~\ref{fig:strip}.
This analysis shows that the most prominent feature is due to electrostatics only: the ``X'' shape corresponds to a gap opened in one bilayer with all carriers hosted on the other bilayer.
On either side of this gap there are band edges with a high DoS.
The toy model of momentum-conserving hybridisation without umklapp produces a vertical stripe near zero filling.
Finally, inclusion of the umklapp processes fully restores the \moire miniband structure, particularly, features at integer fillings of the \moire unit cell. 
We attribute the very low DoS at the centre of the map to a gap opened by the \moire potential.
We note that this is distinct from the spontaneous gap of a few meV observed at the charge neutrality point of ultraclean bilayer graphene~\cite{weitzBrokenSymmetryStatesDoubly2010,velascoTransportSpectroscopySymmetrybroken2012,baoEvidenceSpontaneousGapped2012,freitagSpontaneouslyGappedGround2012}, which is attributed to exchange-driven (Fock) symmetry breaking within each bilayer~\cite{minPseudospinMagnetismGraphene2008,zhangSpontaneousInversionSymmetry2010,zhangSpontaneousQuantumHall2011,nandkishoreDynamicalScreeningExcitonic2010} and is suppressed by displacement fields of order \SI{15}{mV/nm}~\cite{velascoTransportSpectroscopySymmetrybroken2012}; we neglect this Fock contribution here. 

Some of the darkest regions in the DoS maps are related to a vHS at the Fermi level, and are marked by white dotted curves.
Representative Fermi contours along the dotted traces in Fig.~\ref{fig:res2p}(a) are shown in Fig.~\ref{fig:res2p}(e).
The individual bilayer contours are reconstructed by the \moire potential, causing LTs.
Panels (i) and (ii) show LTs of type 1, which reconnect at saddle points and form a connected network spanning across \moire Brillouin zones.
Panels (iii) and (iv) show LTs of type 2 which are isolated to one mBZ without forming a network.
For the antiparallel stacking we do not show these panels because they are very similar to the parallel case.

The longitudinal resistivities normalised by $\tau$ for parallel and antiparallel configurations are shown in Fig.~\ref{fig:res2p}(c) and Fig.~\ref{fig:res2ap}(c), respectively.
The region with the highest longitudinal resistivity is directly related to the bright region (with a very low DoS) in the DoS map.
The dashed lines show positions of the CNPs for bottom and top bilayers in blue and red, as identified in Fig.~\ref{fig:strip}.
The semiclassical Hall coefficients are shown in Fig.~\ref{fig:res2p}(d) and Fig.~\ref{fig:res2ap}(d). The trace of occupied high-symmetry points are also marked by dotted lines.

Corrections to the Hall response from the Berry curvature and orbital magnetic moment of the minibands are independent of $\tau$ and therefore subdominant in the high-mobility regime considered here, except close to miniband edges where the Berry curvature is concentrated.
Preliminary results for this contribution are presented in Appendix~\ref{sec:quantumHall}.

\subsection{Comparison of $q=0$ and finite $q$ terms in the Hartree model}\label{subsec:Hartree discussion}

\begin{figure*}
\includegraphics[width=\textwidth]{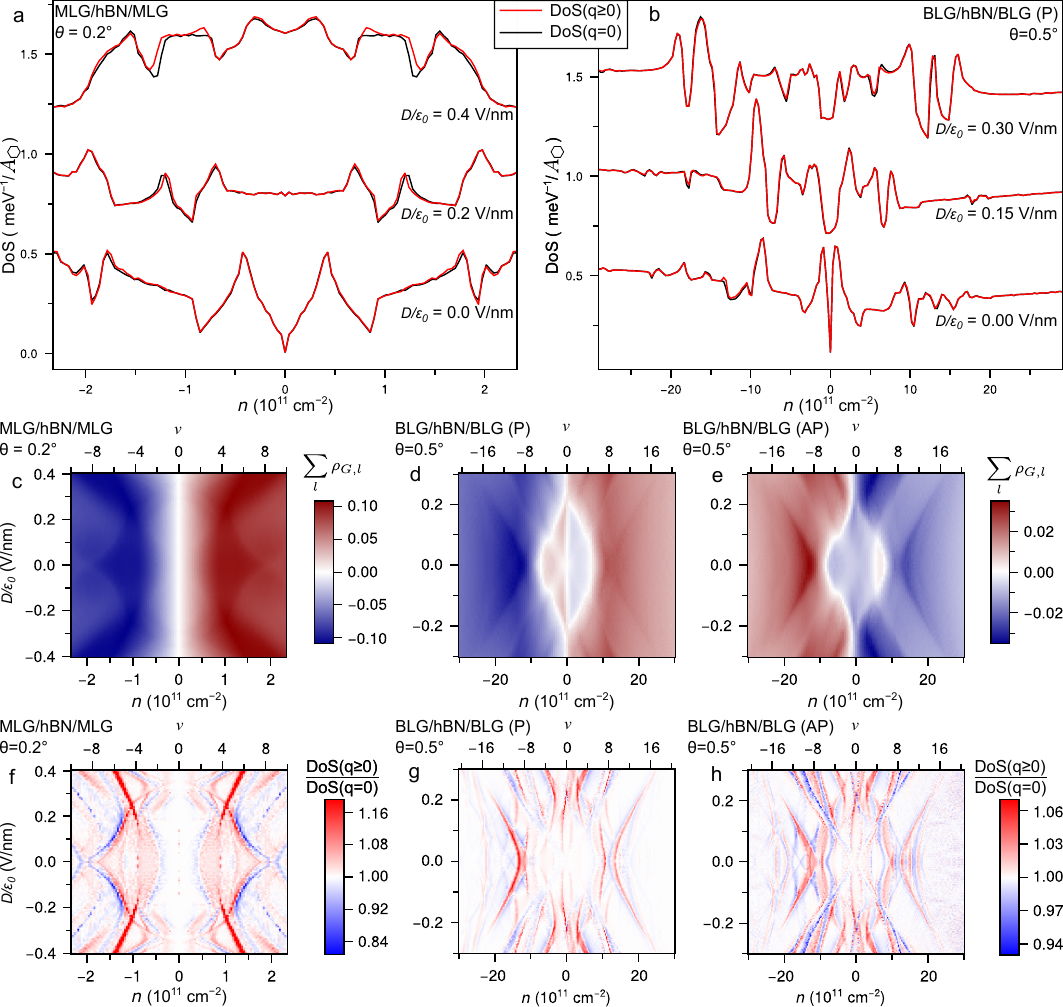}
    \caption{Comparing the significance of finite $q$ Hartree terms to the $q=0$ only model.
    \textbf{(a-b)}: Exemplary line cuts of DoS calculated with (red) and without (black) the finite momentum terms.
    Left: MLG/hBN/MLG, $\theta=0.2\degree$. Right: parallel BLG/hBN/BLG, $\theta=0.5\degree$.
    DoS is plotted against carrier density $n$ for three line cuts at different displacement fields as labelled on the plot.
    The bottom lines are at the numerical DoS as shown, but the middle and top lines are shifted up by 0.5 and 1 units respectively, so they can be distinguished.
    Red lines are the full model ($\mathrm{DoS}(q\geq 0)$); the black lines underneath are the homogeneous part only ($\mathrm{DoS}(q=0)$).
    \textbf{(c-e)}: Maps of the magnitude of the first shell of density fluctuations due to Hartree interactions, for the three systems as labelled.
    \textbf{(f-h)}: Ratio of DoS at the Fermi level from the full model to the $q=0$ only model.
    Overall, a low density variation is induced but it has not much effect on density of states, only some features are slightly shifted.
    }
    \label{fig:q0}
\end{figure*}

In conventional TBG the in-plane charge density fluctuation significantly modifies the miniband structure \cite{guineaElectrostatics2018,ceaBandStructureInsulating2020,ceaElectronicBandStructure2019,ceaElectrostaticInteractionsTwisted2022}.
However, in this system with a reduced \moire coupling, the finite-momentum Hartree potential is much less.
The contribution of the finite-momentum ($\bm{q}\neq0$) Hartree terms is weak throughout the parameter space examined for the BLG/hBN/BLG systems.
Maps of the first-shell density fluctuations (Fig.~\ref{fig:q0}(c-e)) show that this correction is small for the whole range of parameters and the most significant effects occur when the Fermi level is near miniband edges.
Away from those areas the $\bm{q}\neq0$ Hartree potential is almost negligible, so the only important Hartree interaction is the zero-momentum ($\bm{q}=0$) part, the onlayer potentials given in Eq.~\eqref{eq:onlayer}.
The effect of the $\bm{q}\neq 0$ correction on the DoS is shown in Fig.~\ref{fig:q0}(f-h).
The corrections are 10\% at most, with the largest effects occuring when the Fermi level is near miniband edges and van Hove singularities.
Therefore, we have shown that it is not necessary to model the full Hartree interaction to explain features in twist-with-a-spacer systems.
The weakened interlayer coupling reduces the in-plane charge inhomogeneity.
So, in future theoretical descriptions it is sufficient to use the minimal model of homogeneous onlayer potentials for computational efficiency without sacrificing accuracy.

For the MLG/hBN/MLG, we notice that in the majority of the map the DoS was increased by including finite $\bm{q}$ contributions.
This is consistent with the predictions for TBG made in Ref.~\cite{ceaElectronicBandStructure2019}.
The authors showed that the effect of the Hartree interaction is to ``pin'' a higher density of states to the Fermi level for a range of fillings $[-4,4]$.
In general, this effect is also seen in our system, but the magnitude of the pinning is much weaker than in TBG due to the weaker interlayer coupling.
The most prominent enhancement of DoS is at the vHS of type (ii) which presents as the bright red stripes which emanate from $|\nu|=4$ to $|\nu|=6$ at $D/\varepsilon_0=\SI{0.4}{V/nm}$.

Figure~\ref{fig:q0}(a-b) emphasises how small the correction from $q >0$ terms is; we plot line cuts of DoS at different displacement fields for the results of the full model with $q \ge 0$ in red and of the minimal $q=0$ model in black; the lines are almost identical showing the same features although some are slightly shifted.

\section{High-temperature quantum oscillations}

Usually in 2-dimensional systems, the application of a magnetic field causes carriers to follow cyclotron orbits, encircling an area which gives rise to Shubnikov-de Haas (SdH) oscillations and ultimately causes the density of states to discretise into Landau levels \cite{onsagerInterpretation1952,souleStudy1964}.
However, SdH oscillations are sensitive to thermal broadening which mixes phases from orbits of different areas, so must be observed at low temperature.
In marginally twisted MLGs with a spacer, we predict two additional oscillations in magnetic field which are robust at elevated temperatures.
Firstly, we predict a type of SdH oscillation which is tied to the superlattice, when the Fermi level is near an ``open'' LT in the miniband structure.
Secondly, a hybridisation oscillation for low fillings and high displacement fields, where the overlap between electron-like and hole-like Fermi surfaces is an area that is constant for a range of fillings.

\subsection{High-temperature Brown-Zak oscillations}\label{subsec:BZ}

\begin{figure}
    \includegraphics[width=\columnwidth]{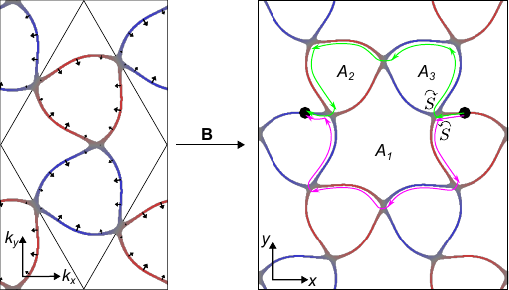}
    \caption{Left: a Fermi contour showing a saddle point associated with the LT and vHS in Fig.~\ref{fig:res1}e(ii), which corresponds to right: real space trajectory, by the application of a magnetic field $\bm{B}$.
    Green and pink lines depict two possible paths which have the same length and enclose an area of one mBZ.
    The saddle points are where interlayer hybridisation permits a carrier to tunnel from one layer to the other or continue in the same layer, splitting a path into two interfering trajectories with scattering amplitudes $\overset{\curvearrowleft}{S}$ and $\overset{\curvearrowright}{S}$.
    }
    \label{fig:BZ}
\end{figure}
\begin{figure*}
    \includegraphics[width=\textwidth]{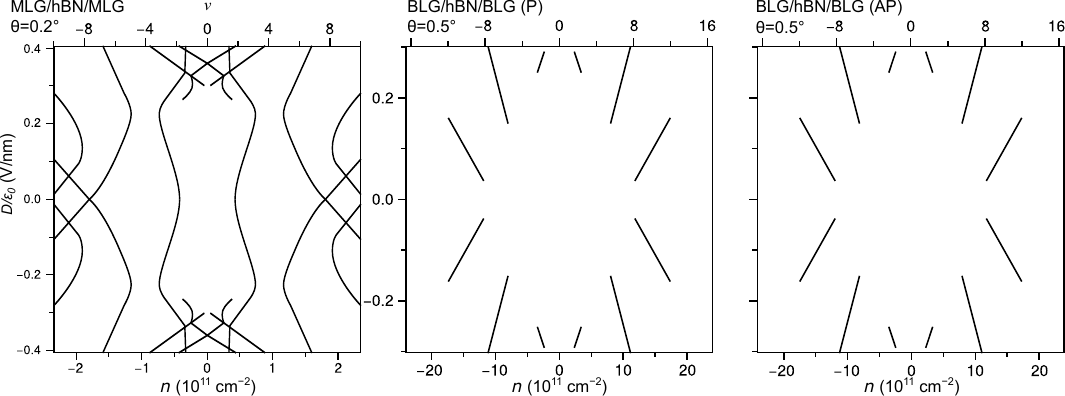}
    \caption{Lines showing LTs for which the high-T Brown-Zak oscillations are predicted.
    Left: MLG/hBN/MLG, $\theta=0.2\degree$. Centre: parallel BLG/hBN/BLG, $\theta=0.5\degree$. Right: antiparallel BLG/hBN/BLG, $\theta=0.5\degree$.}
    \label{fig:vhs}
\end{figure*}

A van Hove singularity (vHS) is a divergence in the DoS which occurs at a saddle point in the dispersion, where the effective mass changes sign and there is a Lifshitz transition.
In a \moire system there are two types of LT; the usual ``closed'' type \cite{varletAnomalous2014}, which is the transition between different local Fermi surface topologies of a single bilayer, and an ``open'' type which is a transition to a connected network of Fermi contours \cite{devriesKagomeQuantumOscillations2024}.
Some examples of both of these are shown above in Figs.~\ref{fig:res1}(e) and \ref{fig:res2p}(e), where the ``open'' type is labelled as LT type 1 and the ``closed'' type is LT type 2.

Under the application of a magnetic field, the Fermi contours are rotated $90\degree$ and scaled to become cyclotron orbits and at the saddle points magnetic breakdown permits a carrier to tunnel between layers and switch to a different trajectory.
Therefore, we can construct two trajectories which take different paths at a vHS.
These enclose an area, so an Aharonov-Bohm phase is accumulated due to the magnetic field, and leads to oscillations of the in-plane longitudinal resistance with period $1/B$.
Specifically, the area enclosed is (an integer multiple of) the \moire unit cell $A_{\varhexagon}$.
These are the kagome-looking networks of ballistic trajectories identified in
Ref.~\cite{devriesKagomeQuantumOscillations2024}.

A typical example is depicted in Fig.~\ref{fig:BZ}, which shows a LT in the MLG/hBN/MLG system where the Dirac cones were filled enough to hybridise.
The $k$-space cut shows almost circular red and blue contours, which are the Dirac cones of top and bottom layers.
In real space, these become cyclotron trajectories on top and bottom layer.
At the saddle points there is a scattering amplitude $\overset{\curvearrowleft}{S}$ to continue on the same trajectory or $\overset{\curvearrowright}{S}$ to switch to the other trajectory, since at these points the carriers are shared between the layers.
This is indicated by the grey colour, which means $\Delta(\bm{k},n)=0$.
Then, at this point a carrier can switch layer.
There are several possible paths which enclose an area multiple of the \moire unit cell, but the one with the shortest lengths is shown.
Importantly, the green and pink trajectories are assembled from the same path segments so they have an identical length $\mathcal{L}$ and the energy dependent phase is the same for both.
The only phase is from the encircled area, which is set by the \moire superlattice and constant for a range of carrier energies, which enables the temperature resilience of this effect.

This interference causes a correction to the longitudinal conductivity
\cite{devriesKagomeQuantumOscillations2024},
\begin{equation}
    \Delta \sigma \sim \frac{e^2}{h}  e^{-2\mathcal{L}/\ell} \int d\epsilon \, \frac{\partial f_0}{\partial \epsilon} |\overset{\curvearrowleft}{S}^6 \overset{\curvearrowright}{S}^4| \times\cos \frac{2\pi\phi_0}{A_{\varhexagon} B},
\end{equation}
where the flux quantum is $\phi_0=h/e$.
The correction is suppressed by scattering by an exponential factor of the ratio of path length $\mathcal{L}$ to the mean free path $\ell$, which determines the ballistic propagation.
The parameters at which there are open LTs that could give rise to this oscillation effect are plotted on the ($n,D$) maps in Fig.~\ref{fig:vhs} for all three systems.

\subsection{High-temperature hybridisation oscillations}\label{subsec:hy}

\subsubsection{Interplay between interlayer tunnelling and magnetoresistance}
Consider a strip of two layers with width $w$ in the $y$-direction and infinite in the $x$-direction. The top ($t$) and bottom ($b$) layers have zero-field longitudinal conductivities $\sigma_0^{t/b}$ and are coupled by vertical conductance density $s$.
Under the effect of a magnetic field the longitudinal conductivities are $\sigma^{t/b}=\sigma_0^{t/b}/(1+\beta_{t/b}^2)$ and Hall conductivities are $\sigma_H^{t/b}=\beta_{t/b}\sigma^{t/b}$ where $\beta_{t/b}=\omega_c^{t/b} \tau$.
The Hall voltage distribution induces potentials $U_{t/b}(x,y) = u_{t/b}(y)-E_x x$ and an interlayer current density $s(u_t-u_b)$, 
and this tunnelling efficiently relaxes the difference between onlayer Hall potentials,
as with magnetoresistance
\cite{alekseevMagnetoresistanceTwoComponentSystems2015,moazzamigudarziUltimateChargeTransport2024}.
Without interlayer tunnelling, the two layers would act as parallel resistors. With interlayer tunnelling switched on, the Hall potential difference $u_t-u_b$ equilibrates at a length scale of $\lambda_\perp=1/\sqrt{s \left(\frac{1}{\sigma^t} + \frac{1}{\sigma^b}\right)}$.
The interlayer current density can be written in terms of the current densities $j^{t/b}_{x,y}(y)$,
\begin{equation}
    \nabla\cdot\bm{j}^t = -\nabla\cdot\bm{j}^b = s(u_t-u_b),
\end{equation}
from which we obtain differential equations
\begin{equation}
\begin{split}
    \partial^2_y(u_t-u_b) &= s(u_t-u_b)\left(\frac{1}{\sigma^t} + \frac{1}{\sigma^b}\right) \equiv \frac{u_t-u_b}{\lambda_\perp^2},\\
    \partial^2_y(u_t+u_b) &= s(u_t-u_b)\left(\frac{1}{\sigma^t} - \frac{1}{\sigma^b}\right),
\end{split}
\end{equation}
where $\lambda_\perp$ is the decay length. 
To find solutions, we set the boundary conditions that the Hall current goes to 0 at the edges of the strip, $y=\pm w/2$.
Then, the longitudinal currents are found by integrating across the strips, $I_x^{t/b}=2\int_0^{w/2} j_x^{t/b}\,\mathrm{d}y$
and the longitudinal resistivity $\rho_\text{tot}=E_xw/(I^t_x+I^b_x)$ is given by
\begin{align}\label{eq:rho_tot}
    \rho_\text{tot}^{-1} = (\sigma^t+\sigma^b)
      &\Bigg[
    1 + \left( \frac{\beta_t \sigma^t +\beta_b\sigma^b}{\sigma^t + \sigma^b} \right)^2 \\
     +& (\beta_t - \beta_b)^2 \frac{\sigma^t \sigma^b}{(\sigma^t+\sigma^b)^2}\tanh\left(\frac{w}{2\lambda_\perp}\right)/\frac{w}{2\lambda_\perp}
    \Bigg].\notag
\end{align}
In the case of $n_\text{tot}=0$, then $\sigma^t=\sigma^b\equiv\sigma$ and $\beta_t=-\beta_b\equiv\beta$ and the resistivity simplifies to 
\begin{equation}
\rho_\text{tot}^{-1}=2\sigma\qty[1+\beta^2 \tanh\left(\frac{w}{2\lambda_\perp}\right)/\frac{w}{2\lambda_\perp}]. 
\end{equation}
In the case of no hybridisation and therefore no interlayer tunnelling $s\to 0$, the decay length is long $\lambda_\perp\gg w$ and the expression simplifies to the sum of two strips in parallel, $\rho_\text{tot}^{-1}=2\sigma_0$.
On the other hand, for high hybridisation and short decay length $\lambda_\perp \ll w$, the resistance is enhanced as $\rho_\text{tot}^{-1} = 2\sigma\left(1+\beta^2\sqrt{\frac{2\sigma}{sw^2}}\right)$.
We are interested in the regime of high mobility ($\beta \gg 1$), since this is where the effect is strongest, but also because such high quality samples are required to see \moire.

In the case of a small displacement field $D\approx0$, the two layers are almost equivalent,
although $D$ splits the layer potentials and densities, $u\propto D$ and $n_{t,b}=n_\text{tot}/2\pm \mu u/(\pi\hbar^2v^2)$, so $\sigma^{t/b}\equiv\sigma(1\pm\frac{u}{2\mu})$ and $\beta_{t/b}\equiv\beta(1\mp\frac{u}{2\mu})$.
Substitution into Eq.~\eqref{eq:rho_tot} gives
\begin{align}
    \rho_{\mathrm{tot}}^{-1}=2\sigma_0
    -\sigma\beta^2\frac{u^2}{2\mu^2}\qty[1-\tanh\left(\frac{w}{2\lambda_\perp}\right)/\frac{w}{2\lambda_\perp}],
    \label{eq:rho_smallD}
\end{align}
and when $D=0$ the resistivity reduces to $\rho_{\mathrm{tot}}^{-1}=2\sigma_0$.
For high hybridisation and high mobility, we obtain $\rho_\text{tot}^{-1} \approx 2\sigma_0\left(1-\left(\frac{u}{2\mu}\right)^2\left(1-\frac{1}{\beta w}\sqrt{\frac{\sigma_0}{2s}}\right)\right)$.

The hybridisation $\Pi$ is not measured directly, but it affects in-plane and vertical transport through the vertical conductance density $s$ that couples the two graphene sheets, which is related 
to the hybridisation of wavefunctions on top and bottom layers, $\Pi$, and the DoS in top and bottom layers by \cite{simmonsGeneralizedFormulaElectric1963,feenstraSingleparticleTunnelingDoped2012,laneTwistcontrolledResonantTunnelling2015,katkovReviewArticleTunnelingbased2017,feenstraMagnitudeCurrent2D2018}
\begin{equation}
    s \propto \frac{e^2}{h} A_{\varhexagon} \text{DoS}_t\,\text{DoS}_b\,\gamma^2\Pi^2.
\end{equation}
An oscillation of $\Pi(B)$, and hence of $s(B)$ and $\lambda_\perp(B)$, therefore appears in the in-plane magnetoresistance $\rho_\text{tot}$.

\subsubsection{Hybridisation oscillations effect on magnetoresistance}

We also predict high-temperature hybridisation oscillations in a magnetic field due to the overlap between Dirac cones of the two layers.
The different conditions by which the two cones can overlap are depicted in Fig.~\ref{fig:res1}(f) with marks corresponding to regions of the hybridisation map in Fig.~\ref{fig:res1}(b).
The overlap that produces these oscillations is fixed by two parameters: the Fermi level $\mu$ and the layer offset $u\equiv\epsilon_t-\epsilon_b$ set by the displacement field $D$.
The Dirac cones of the two layers are separated in momentum by $\Delta K$.
Therefore, there is a critical value of displacement field $D^*$ which causes the onlayer potential difference $|u|=\hbar v\Delta K$ so that the two cones coincide along an axis at all fillings and there is resonant tunnelling \cite{britnellResonantTunnellingNegative2013,mishchenkoTwistcontrolledResonantTunnelling2014,laneTwistcontrolledResonantTunnelling2015}.
In order for there to be an overlap between opposite carrier type Fermi surfaces, the magnitude of displacement field must be higher than this critical value.

\begin{figure*}
    \includegraphics[width=\textwidth]{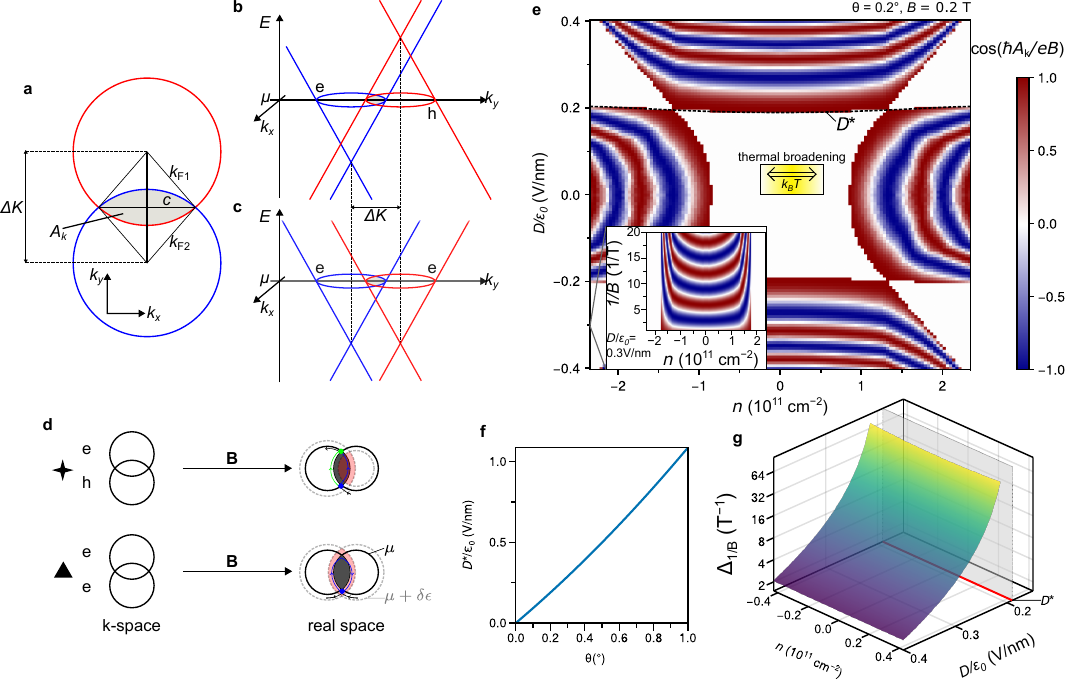}
    \caption{Magnetic oscillations from intersecting Fermi cones.
    \textbf{(a)}: Fermi cut showing geometry of the circles with Fermi wavevectors $k_{Fi}$, displaced by $\Delta K$, which intersect in the grey shaded area $A_k$ and the chord between the points of intersection has length $2c$.
    \textbf{(b)} and \textbf{(c)}: dispersions showing the two types of intersection,
    where carriers in each layer are of opposite type (electrons (e) and holes (h)) or the same type, respectively.
    \textbf{(d)}: a magnetic field sets up cyclotron orbits by rotating the Fermi contours and scaling by $\lambda_B^2$.
    The two regimes correspond to the shapes on Fig.~\ref{fig:res1}b and (b)/(c) in this figure.
    \textbf{(e)}: oscillating factor of the interlayer hybridisation due to overlap area between Dirac cones, $\cos(\hbar A_k/eB)$ in Eq.~\eqref{eq:PiB}, which modulates the in-plane magnetoresistance through the interlayer conductance $s$, for the MLG/hBN/MLG system at $\theta=0.2\degree$ with $B=\SI{200}{mT}$, plotted on a map of displacement field against filling (shown only where the Fermi contours overlap, $A_k\neq0$).
    Thermal broadening of $k_B T$ is equivalent to an averaging over a small range of fillings, as indicated by the arrow in the yellow box.
    Inset: the same quantity plotted on a map of $1/B$ against filling, for the line $D/\varepsilon_0=\SI{0.3}{V/nm}$.
    \textbf{(f)}: dependence of the critical displacement field $D^*$ on twist angle $\theta$ (at $n=0$).
    \textbf{(g)}: the oscillation periodicity in $1/B$ for the temperature resilient part at low fillings and displacement field above $D^*$.
    }
    \label{fig:qmt}
\end{figure*}

We neglect the higher-order Hartree corrections and the \moire potential and just use the onlayer potentials produced by the self-consistent algorithm for each ($n$, $D$) point on the maps.
These two parameters define the density hosted in each layer $i$: $n_i=(\mu\pm u/2)^2/(\pi\hbar^2v^2)\ \text{sign}(\mu\pm u/2)$ and $n_t+n_b=n$.
The intersection of the two Fermi contours has area $A_k$ and the two arcs have length $L_{k,i}$, which can be found using some geometry.
In general, we have
\begin{equation}
\begin{split}
  A_k &= \pi|n_t| \arcsin{\frac{c}{\sqrt{\pi|n_{t}|}}} + \pi|n_b| \arcsin{\frac{c}{\sqrt{\pi|n_{b}|}}} - c \Delta K,\\
  L_{k,i} &= 2\sqrt{\pi|n_i|} \arcsin{\frac{c}{\sqrt{\pi|n_{i}|}}},
\label{eq:ak}
\end{split}
\end{equation}
where
\begin{equation*}
  c = \frac{\Delta K}{2} \sqrt{\left[\left(\frac{u}{\hbar v \Delta K}\right)^2 - 1\right] \left[ 1 - \left(\frac{2\mu}{\hbar v \Delta K}\right)^2\right]}
\end{equation*}
is the distance along the $\mu-\gamma$ line at which the Fermi contours cross, equal to half the chord length.
The area formula can be understood as the area of each circular sector minus the triangular portion as shown in Fig.~\ref{fig:qmt}a, which shows the Fermi cuts with the Fermi wavevector of each layer $k_{Fi}$, which is related to the Fermi level by $\hbar v k_{Fi}= |\mu\pm u/2|$.
The meaning of the path length difference depends on the carrier trajectory: either the difference between each arc (for e/h) or the total length of the intersection (for e/e and h/h).

In the presence of a magnetic field, the real-space cyclotron orbits follow the Fermi contours rotated by $90\degree$ and scaled by $\lambda_B^2=\frac{\hbar}{eB}$, as shown in Fig.~\ref{fig:qmt}d.
Therefore, areas are scaled by $\lambda_B^4$.
The Aharonov-Bohm phase difference $\Delta\phi$ between paths which circle the real space area $A=\lambda_B^4 A_k$ causes hybridisation oscillations as
\begin{equation}
\begin{split}
    \Delta\phi &= 2\pi \Phi/\phi_0 = 2\pi BA \frac{e}{h} = \frac{\hbar}{eB}A_k,\\
    \Pi &\to (1+\alpha\cos(\hbar A_k/eB))\Pi \label{eq:PiB}
\end{split}
\end{equation}
with $\alpha < 1$.
This hybridisation oscillation affects the in-plane conductivity as explained above, leading to oscillations of in-plane magneto-transport.
These oscillations are illustrated in Fig.~\ref{fig:qmt}e for $B=\SI{200}{mT}$, where the parts lacking resonant hybridisation between Dirac cones are shown in white, and 
we identify 4 regions of interest where Fermi lines overlap.
These edges of these regions are marked by the same lines as in Fig.~\ref{fig:res1}b which marked where Fermi lines touch and the critical displacement field which causes cones to touch along a line.

The regions on the top and bottom are where the overlapping cones have opposite carrier type, like in Fig.~\ref{fig:qmt}b, and the regions on the side are where the overlapping cones have the same carrier type, like in Fig.~\ref{fig:qmt}c.
To consider the behaviour at elevated temperature, thermal broadening enables mixing of states above and below the Fermi level to affect transport characteristics.
This is akin to mixing different fillings, or averaging over the horizontal axis, by taking $\mu\to\mu+\delta\epsilon$.
Therefore, it is clear that the top/bottom regions of e/h overlap, which are almost constant for a wide range of fillings, will be more robust than the side regions of e/e or h/h overlap, which vary more rapidly.
This is indicated by the arrow and yellow box depicting thermal broadening in panel (e).

To confirm this analysis and gain an intuition about the dependence of the area on changes in the Fermi level,
we can take Taylor approximations for small $c$, which describe the behaviour at the innermost edges of those regions, where the overlap is very small.
Specifically, in regimes where $c\ll k_{Ft}, k_{Fb}$ so that the arcsin can be expanded for small $x$ as $\arcsin x \approx x + \frac{x^3}{6}$.
This condition is held in two types of regime.
Firstly, when $\mu \approx 0$ and $|u| \approx \hbar v\Delta K$;
which is one hole- and one electron-type, with a high displacement field and filling around 0.
The second type of regime is around $u\approx 0$ and $2|\mu| \approx \hbar v\Delta K$, which is two electron (or two hole) like circles.
In each case we are interested in the dependence on $\mu$ because small changes in $\mu$ are like thermal broadening.
In the first case, small $\mu$ and opposite sign densities in each layer:
\begin{align*}
  A_k &\approx \frac{2\sqrt{2}}{3} \Delta K^2 \left(\frac{u}{\hbar v\Delta K}-1\right)^{3/2} \left[1 - 2\frac{\mu^2}{(\hbar v \Delta K)^2} \right].
\end{align*}
In the second case, small $u$ and same sign of densities in each layer:
\begin{align*}
  A_k &\approx \frac{2\sqrt{2}}{3} \Delta K^2 \left(\frac{2|\mu|}{\hbar v\Delta K}-1\right)^{3/2} \left[1 - \frac{u^2}{2(\hbar v \Delta K)^2} \right].
\end{align*}
This analysis shows that the e/h type overlap has a dependence on $\delta\epsilon^2$, but the e/e and h/h type overlap has a dependence on $\delta\epsilon^{3/2}$.
For e/h overlap there is also a difference in path lengths which adds some phase interference that is asymmetric in filling factor: $\phi_L = \hbar/eB (k_{Ft}L_{k,t} - k_{Fb}L_{k,b})$.
For e/e or h/h, the interfering particle goes around the whole contour so it is the total length of the arcs, not their difference, which gives a phase difference, which is a much faster oscillation.

The periodicity in $1/B$ is $\Delta_{1/B} = \frac{2\pi e}{\hbar A_k}$ with $A_k$ given by Eq.~\eqref{eq:ak}.
This formula requires the output of the self-consistent Hartree analysis to convert parameters $n,D$ to energies $\mu, u$.
The result, using the self-consistent Hartree, is plotted in Fig.~\ref{fig:qmt}(g).
This hybridisation oscillation is periodic in $1/B$ and predicted to occur at low fillings (roughly, $|\nu|<4$) and displacement fields greater than the critical resonant field $D^*$, which is defined as the field required to cause $u = \hbar v \Delta K$.
This effect does not depend on the \moire superlattice, only on hybridisation between offset Dirac cones.
A linear formula cannot be written because screening requires the self-consistent algorithm to be used to find $D^*$.
However, a lower bound can be found by neglecting screening,
and the behaviour of the periodicity in $1/B$ near the threshold reads as 
\begin{equation}
\begin{split}
D^* &\approx \hbar v K\theta \frac{\varepsilon_0}{e}\left(\frac{d_\text{G}}{\varepsilon_z^\text{G}} + \frac{d_\text{hBN}}{\varepsilon_z^\text{hBN}}\right)^{-1},\\
\Delta_{1/B} &\sim \theta^{-2} \left(\frac{D}{D^*}-1\right)^{-3/2},\\
\Pi &\to \Pi\left(1 +\alpha\cos\left[\frac{2\sqrt{2}\hbar\Delta K^2}{3eB}\left(\frac{D}{D^*}-1\right)^{3/2}\right] \right).
\end{split}
\end{equation}
The precise relationship between $D^*$ and $\theta$ accounting for screening is plotted in Fig.~\ref{fig:qmt}(f).
The periodicity in $1/B$ for $\theta=0.2\degree$ and low fillings is plotted in Fig.~\ref{fig:qmt}(g).
For the oscillations to occur, there must be some finite overlap hence the requirement for $D>D^*$.
As the displacement field reduces and approaches $D^*$, the area goes to 0 so the periodicity diverges.
This means that the oscillation becomes very slow near the threshold, so to observe the effect at reasonable field strengths we must be a little above $D^*$.
Therefore, the plot is limited so that some details are visible, and the plane $D=D^*$ is shown in grey with a red line at the bottom.

These hybridisation oscillations are also expected to occur in BLG/hBN/BLG.
However, additional study is required to understand the extra complexity and symmetry considerations in bilayer dispersions.

\section{Conclusion}

Small-angle-twisted graphene devices with a monolayer spacer exhibit a range of features due to the interplay between \moire superlattice effects and interlayer hybridisation.
We have modelled 3 such systems: MLG/hBN/MLG, a twisted bilayer graphene with an hBN spacer; and BLG/hBN/BLG, a twisted double bilayer graphene with an hBN spacer, in two configurations (AB/AB and AB/BA).
The monolayer \textbf{spacer reduces the tunnelling amplitude} between the graphenes to $\gamma\approx\SI{7}{meV}$, about fifteen times weaker than in conventional directly-contacting TBG or TDBG.
At the same time, the greater layer separation enhances the effect of the displacement field.

For each system we have calculated the dispersions while varying an external displacement field and total carrier density, and from these we traced features including miniband edges, charge neutrality points, resonant tunnelling conditions, and van Hove singularities.
We also modelled semiclassical transport characteristics and noted parametric intervals where the effect of such features is notable.
The distinction between the two configurations of double bilayers is reduced due to the weaker coupling strength, in contrast to conventional TDBG.
We predict features due to \moire minibands, but there is more to be understood to completely describe conduction accounting for impurities, edge effects, and high magnetic fields.

We compared self-consistent Hartree calculations with ($q \ge 0$) and without ($q=0$) an in-plane periodic charge redistribution.
The weakened interlayer coupling means that the essential miniband structure and hybridisation trends are well accounted for by the minimal electrostatic model, which is \textbf{computationally more efficient} yet quantitatively useful as a model for graphene/hBN/graphene \moire systems over a wide range of twist angles, displacement fields, and fillings.

The \moire potential causes Lifshitz transitions which connect Fermi lines to form a \textbf{network across \moire Brillouin zones}.
Under the application of a magnetic field, these Fermi lines become semiclassical trajectories, and at the van Hove singularities where bands touch, interlayer hybridisation permits a carrier to tunnel from one layer to another.
Pairs of trajectories which part and rejoin at these crossings enclose an area set by the \moire unit cell which contributes a magnetic phase, resulting in $1/B$-periodic magnetoresistance oscillations of Brown-Zak type but persistent to weak magnetic fields and which are expected to survive to a higher temperature than conventional Shubnikov-de Haas oscillations.

A strong displacement field can offset Dirac cones of the graphenes so that there is an overlap between Fermi surfaces of opposite carrier types at low fillings.
At the intersection points carriers can tunnel between layers, which provides an opportunity for interference enclosing an overlap area that is approximately constant across a range of fillings and therefore robust to thermal broadening of the Fermi step.
This interference generates a \textbf{peculiar type of magneto-oscillations of interlayer hybridisation} which leads to measureable $1/B$-periodic quantum oscillations of resistivity in high-mobility devices and persistent to high temperature.

The \moire effectiveness is controlled by the twist angle.
At large $\theta \ge 1\degree$, the \moire potential is too weak to support miniband formation, but interlayer hybridisation with momentum conservation is still efficient.
The hybridisation oscillations, which do not rely on the superlattice, are visible at low fillings and displacement fields above the critical value $D^*$ which is set by the twist.
At intermediate $\theta\sim 0.2-0.7\degree$, the \moire potential creates minibands and causes LTs which host Brown-Zak oscillations, for a range of external displacement fields.
At extremely small $\theta\le0.1\degree$, \moire dominates the behaviour.
Therefore, we particularly recommend exploring twist angles $\theta\sim0.5\degree$ as a unique weakly-coupled \moire system in which the hybridisation between dispersions of each layer can be tuned to see Brown-Zak or hybridisation oscillations.

\section*{Acknowledgments}
X.L., H.F. and Y.P. are supported by the Research Grants Council of Hong Kong (CityU 11312825, C7012-21G, and C7015-24G) and City University of Hong Kong (Project No. 9610428). B.F. acknowledges Quantum funding in EPSRC DTP grant EP/W524347/1. V.I.F. acknowledges support from the Mat4QE project funded by the National Physical Laboratory.

X.L. edited some texts in the manuscript with the assistance of an LLM. 
All content was reviewed and approved by the authors, who take full responsibility for the manuscript.

\section*{Data Availability}
There are no publicly available research data or software
supporting this manuscript. Requests for further information or data should be sent to the authors.


\appendix
\counterwithin{figure}{section}
\section{Methods}
\label{sec:methods}
\subsection{Details of numerical calculations}

\begin{table}[b]
    \caption{Values of parameters used in the numerical calculations 
\cite{slizovskiyOutofPlaneDielectricSusceptibility2021,laturia2018dielectric,pierret2022dielectric,wehlingStrengthEffectiveCoulomb2011,dresselhausIntercalationCompoundsGraphite2002,PhysRevB.99.235406}.}
\label{tab:params}
    \begin{tabular}{lrlr}
    \hline
      Parameter & Value & Parameter & Value\\
      \hline
        $\varepsilon_z^\text{G}$ & 2.6 & $v$ & \SI{1.023e6}{m/s} \\
      $\varepsilon_z^\text{hBN}$ & 3.6 & $v_3$ & \SI{1.230e5}{m/s} \\
      $\varepsilon_\parallel$ & 7.0 & $v_4$ & \SI{4.530e4}{m/s}\\
        $a$ & \SI{0.246}{nm} & $\gamma_1$ & \SI{381}{meV} \\
      $d_\text{G}$ & \SI{0.335}{nm} & $\Delta'$ & \SI{25}{meV} \\
      $d_\text{hBN}$ & \SI{0.333}{nm} & & \\
      \hline
    \end{tabular}
\end{table}

In numerical calculations, we used the values for parameters as shown in Table~\ref{tab:params}.
We note that the in-plane dielectric constant of the environment, $\varepsilon_\parallel$, varies over a wide range of values in the literature, around 4-10 for TBG systems \cite{ceaElectronicBandStructure2019}.
For the continuum \moire model, the basis is limited by a cutoff $\bm{G}_m=3|G|$ such that only \moire Bragg vectors with $|\bm{G}|\le|\bm{G}_m|$ are included.

The numerical calculations presented in this work were performed using the Julia programming language \cite{Julia-2017} and figures were produced with the Makie.jl package \cite{DanischKrumbiegel2021}.

\subsection{Conductivity calculation}\label{Conductivity calculation}

Consider an electron with charge \(q = -e\) (\(e > 0\)) moving under the influence of an electric field \(\mathbf{E} = (E_x, E_y, 0)\) and a weak magnetic field \(\mathbf{B} = (0, 0, B)\). Neglecting the Berry curvature and orbital magnetic moment of the minibands (their contribution is discussed in Appendix~\ref{sec:quantumHall}), the semiclassical equations of motion read
\begin{align}
    \dot{\mathbf{r}} &= \mathbf{v}, \label{eq:r_dot} \\
    \hbar \dot{\mathbf{k}} &= -e\mathbf{E} - e(\dot{\mathbf{r}} \times \mathbf{B}), \label{eq:k_dot}
\end{align}
with the group velocity \(\mathbf{v} = \frac{1}{\hbar}\nabla_{\mathbf{k}}\epsilon(\mathbf{k})\).

We consider the system to have a long scattering time $\tau$ which characterises any scattering processes.
Under the relaxation-time approximation the Boltzmann equation is given by 
\begin{equation}
    \dot{\mathbf{k}}\cdot\nabla_{\mathbf{k}}f = -\frac{f_{1}}{\tau},
\end{equation}
where \(f = f_{0}+f_{1}\). Here \(f_{0}\) is the equilibrium Fermi--Dirac distribution and \(f_{1}\) is the non-equilibrium deviation. We assume a constant relaxation time \(\tau\) independent of wavevector.

Collecting terms up to first order in both \(\mathbf{E}\) and \(\mathbf{B}\), one obtains the conductivity tensor components, in one valley and spin \cite{ongGeometric1991},
\begin{equation}
\begin{aligned}
    \sigma_{\alpha\alpha}
    &= e^{2}\tau\int\frac{\mathrm{d}^{2}\boldsymbol{k}}{(2\pi)^{2}}\,
    v_{\alpha}^{2}
    \Biggl(-\frac{\partial f_{0}}{\partial\epsilon}\Biggr),\\
    \sigma_{\alpha\beta}
    &= \frac{e^{3}\tau^{2}B}{\hbar}\int\frac{\mathrm{d}^{2}\boldsymbol{k}}{(2\pi)^{2}}\,
    v_{\alpha}
    \Biggl(v_{y}\frac{\partial}{\partial k_{x}}-v_{x}\frac{\partial}{\partial k_{y}}\Biggr)
    v_{\beta}
    \Biggl(-\frac{\partial f_{0}}{\partial\epsilon}\Biggr).
\end{aligned}
\end{equation}
Summing over valleys and including spin degeneracy, we have the longitudinal part $\sigma = 2e^2\tau \int \frac{\mathrm{d}^2 \bm{k}}{(2\pi)^2} |v|^2 \left(-\frac{\partial f_{0}}{\partial\epsilon}\right)$
and the Hall part 
\begin{widetext}
\begin{equation}
\sigma_H = 2e^3\tau^2 B \int \frac{\mathrm{d}^2 \bm{k}}{(2\pi)^2} \left[v_x^2 (M^{-1})_{yy} + v_y^2 (M^{-1})_{xx} - 2 v_x v_y (M^{-1})_{xy}\right] \left(-\frac{\partial f_{0}}{\partial\epsilon}\right),
\end{equation}
\end{widetext}
where $(M^{-1})_{ij} = \hbar^{-2}\,\partial^2\epsilon/\partial k_i\partial k_j$
 is the inverse effective-mass tensor.

\subsection{Toy model of interlayer hybridisation}
\label{subsec:toy}
\begin{figure*}
    \includegraphics[width=\linewidth]{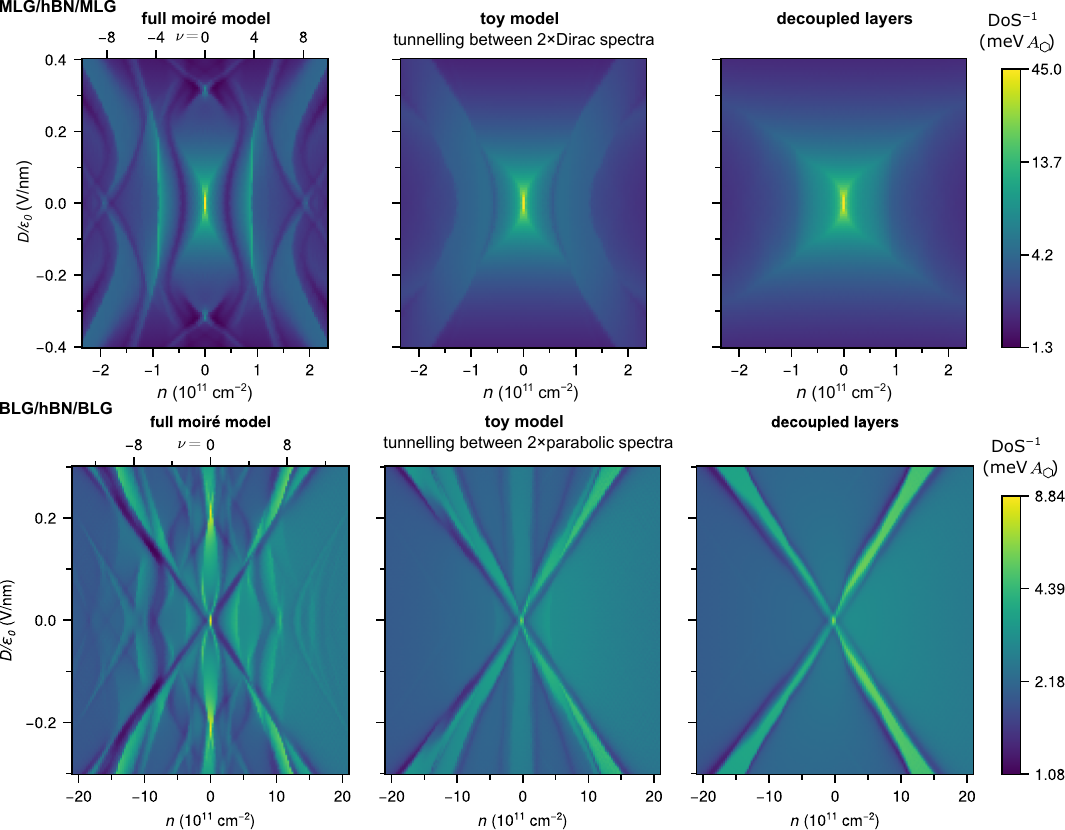}
    \caption{Identifying the origin of features in the inverse DoS maps by inspecting models with some processes turned off.
    Top row: MLG/hBN/MLG at $\theta=0.2\degree$.
    Bottom row: parallel BLG/hBN/BLG, $\theta=0.5\degree$ (antiparallel is indistinguishable).
    Left: full \moire model.
    Middle: toy model interlayer hybridisation without umklapp processes.
    Right: no tunnelling between the graphene bilayers, only electrostatic interactions.
    The colour map is the same for all panels in each row.
    }
    \label{fig:strip}
\end{figure*}

To identify which effects of the model are responsible for various features of the bandstructure and DoS, we compared simpler models with certain tunnelling processes turned off.
Starting with the full \moire model, we distinguish tunnelling processes that conserve momentum between the layers from umklapp processes associated with \moire minibands, which involve tunnelling and scattering by a \moire Bragg vector.

We use a toy model without umklapp scattering which corresponds to two tunnelling-coupled 2D Dirac systems offset in momentum by $\Delta K$.
Such a model is obtained by taking only the $j=0$ term in the tunnelling matrix, main Eq.~\eqref{eq:T}.
Going further, we can completely decouple the layers by setting $\gamma$ to 0, so that the layers have independent dispersions and only interact electrostatically.
The results of these models are plotted in Fig.~\ref{fig:strip} for monolayers and bilayers (there is no distinction between parallel and antiparallel stackings).
Comparing the three successive maps allows every prominent feature in the complete calculation to be assigned unambiguously to electrostatics, hybridisation or umklapp scattering.

\subsection{Traces of occupation of high-symmetry points}
\begin{figure*}
   \includegraphics[width=\linewidth]{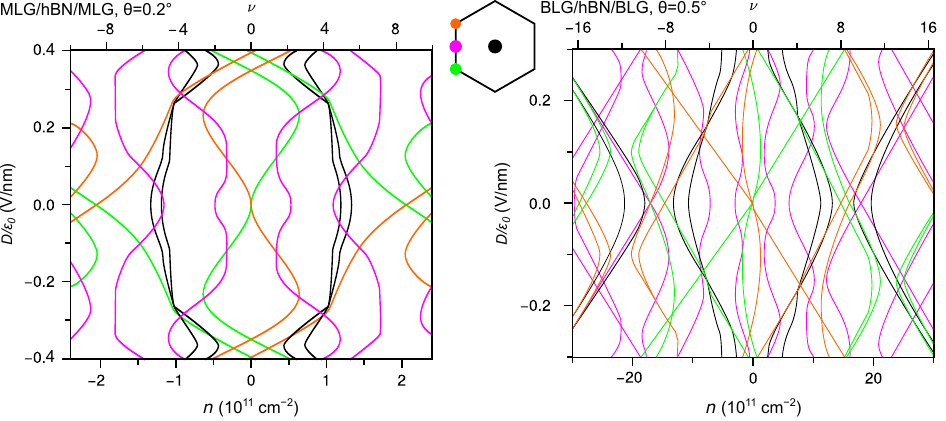} 
   \caption{The coloured lines show where states at high-symmetry points in $k$ space are occupied at the Fermi level.
   Left: MLG/hBN/MLG, $\theta=0.2\degree$. Right: BLG/hBN/BLG, $\theta=0.5\degree$, parallel stacking (antiparallel is indistinguishable).
   The inset shows a \moire Brillouin zone with those high symmetry points labelled with coloured dots.
   }
   \label{fig:hsw}
\end{figure*}

High-symmetry points of the mBZ are usually occupied at band edges or saddle points.
Figure~\ref{fig:hsw} traces lines where each high symmetry point is occupied at the Fermi level.
Some of these lines are overlaid in Fig.~\ref{fig:res1}(d) for MLG/hBN/MLG and in Fig.~\ref{fig:res2p}(d), Fig.~\ref{fig:res2ap}(d) for parallel and antiparallel BLG/hBN/BLG, respectively, but all are shown here.

\section{Results for other twist angles}
\label{sec:other}
\begin{figure*}
    \includegraphics[width=\textwidth]{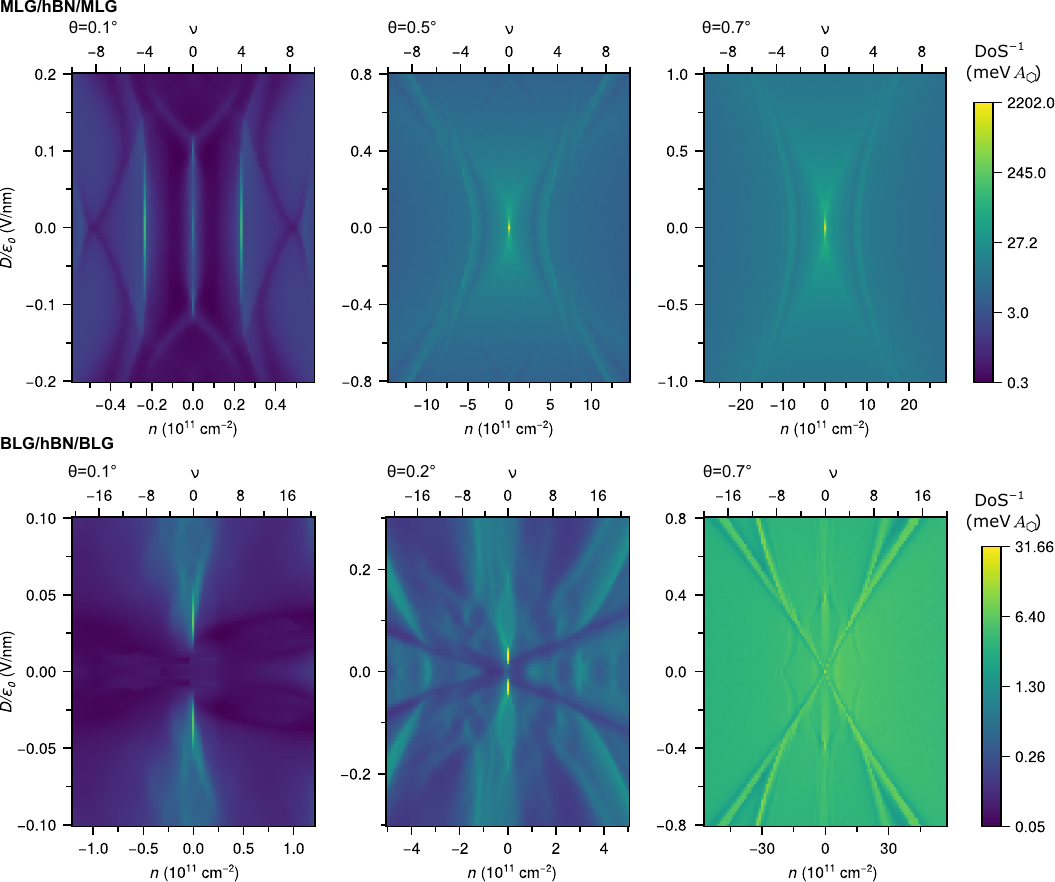}
    \caption{Inverse DoS maps for different twist angles. Top: MLG/hBN/MLG: from left to right, $\theta=0.1\degree, 0.5\degree, 0.7\degree$.
    Bottom: parallel BLG/hBN/BLG: from left to right, $\theta=0.1\degree, 0.2\degree, 0.7\degree$.
    Note that the axis limits change for each plot.
    The colour scale is the same for each row.
    }
    \label{fig:thetas}
\end{figure*}

Figure~\ref{fig:thetas} shows maps of inverse DoS for other twist angles in the MLG/hBN/MLG and parallel BLG/hBN/BLG systems.
In each plot the range of densities has been scaled to be normalised to the same fillings of the \moire unit cell, and the displacement field range has been scaled to have similar effects for the different momentum mismatches.

At $\theta=0.1\degree$, \moire is very strong and the maps show prominent features at integer $\nu$ which remain up to higher displacement fields.
At $\theta=0.7\degree$, we see the influence of interlayer hybridisation but only a few signatures  of \moire.
So, there is an optimal range of twist angles for which the \moire strength is comparable to interlayer coupling and electrostatic effects, such that the system behaviour is highly tunable. 

\section{Berry-curvature and orbital-moment contribution to the Hall conductivity}
\label{sec:quantumHall}
Here we present preliminary results for the contribution of the Berry curvature $\Omega_z(\bm{k})$ and orbital magnetic moment $m_z(\bm{k})$ of the minibands to the Hall conductivity of BLG/hBN/BLG, which is not included in the semiclassical analysis of Sec.~\ref{sec:blg}.
In a weak magnetic field the band energy is corrected as $\epsilon(\bm{k}) = \epsilon^{0}(\bm{k}) - m_z(\bm{k})B$, and the velocity in Eq.~\eqref{eq:r_dot} acquires the anomalous term, $\dot{\mathbf{r}} = \mathbf{v} - \dot{\mathbf{k}}\times\boldsymbol{\Omega}$, with $\mathbf{v} = \hbar^{-1}\nabla_{\mathbf{k}}\epsilon^{0}(\mathbf{k})$.
In addition to the Lorentz term derived in Appendix~\ref{Conductivity calculation}, the Hall conductivity then contains an intrinsic part, independent of $\tau$, which for a single valley and spin reads
\begin{equation}
\sigma_{xy}^{\mathrm{int}} = -\frac{e^2}{\hbar}\int \frac{\mathrm{d}^2 \bm{k}}{(2\pi)^2}\, \tilde{\Omega}_z(\bm{k})\, f_0\bigl(\tilde{\epsilon}(\bm{k})\bigr),
\label{eq:sint}
\end{equation}
with $\tilde{\epsilon} = \epsilon^{0} - m_z B$ and $\tilde{\Omega}_z = \Omega_z + \Omega'_z$.
Here $\Omega'_z = (\nabla_{\bm{k}}\times\bm{a}')_z$ is the field-induced correction to the Berry curvature, which originates from the positional shift $\bm{a}'\propto B$ of the wave packet due to field-induced interband mixing~\cite{gaoFieldInducedPositional2014}.
The part of Eq.~\eqref{eq:sint} linear in $B$, summed over valleys and spin, contains two contributions of the same order,
\begin{equation}
\sigma_H^{(1)} = \sigma_H^{B} - 4\frac{e^2}{\hbar}\int \frac{\mathrm{d}^2 \bm{k}}{(2\pi)^2}\, \Omega'_z f_0,
\label{eq:sH1}
\end{equation}
where
\begin{equation}
\sigma_H^{B} = -4\frac{e^2}{\hbar}B \int \frac{\mathrm{d}^2 \bm{k}}{(2\pi)^2}\, \Omega_z m_z \left(-\frac{\partial f_{0}}{\partial\epsilon}\right)
\label{eq:sHB}
\end{equation}
arises from the orbital-moment correction to the band energy, and the factor of 4 accounts for valley and spin degeneracy.
Both terms are linear in $B$ and independent of $\tau$~\cite{gaoFieldInducedPositional2014}.
In this work we evaluate only $\sigma_H^{B}$; the field-induced Berry-curvature term in Eq.~\eqref{eq:sH1} is not included.
In our model of MLG/hBN/MLG the minibands carry no Berry curvature, so this contribution is absent there.

\begin{figure}
    \includegraphics[width=\columnwidth]{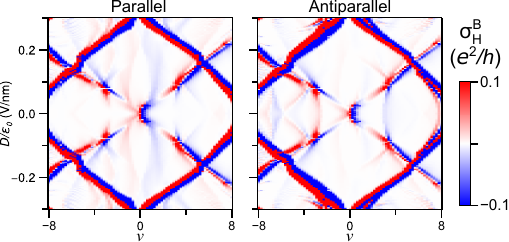}
    \caption{Preliminary results for the Berry-curvature and orbital-moment contribution to the Hall conductivity, $\sigma_H^{B}$ of Eq.~\eqref{eq:sHB}, in BLG/hBN/BLG at $\theta=0.5\degree$, for parallel (left) and antiparallel (right) stacking, calculated at $B=\SI{50}{mT}$. The colour scale is saturated at $\pm0.1\,e^2/h$.}
    \label{fig:sH}
\end{figure}

Figure~\ref{fig:sH} shows $\sigma_H^{B}$ for parallel and antiparallel BLG/hBN/BLG samples at $\theta=0.5\degree$.
It is concentrated along the miniband edges and Lifshitz transitions, where $\Omega_z$ and $m_z$ are largest.
Since $\sigma_H^{B}$ does not depend on $\tau$, while the Lorentz term scales as $\tau^2 B$, the two cannot be compared without choosing a value of $\tau$; in clean samples the Lorentz term dominates away from the band edges, whereas a reduced scattering time, due to impurities or inhomogeneities in the \moire potential, makes $\sigma_H^{B}$ more significant near them.

\bibliography{ref}

\end{document}